\documentclass[12pt,a4paper]{article}

\usepackage{graphicx}

\input{tcilatex}
\begin{document}

\setcounter{page}{0} \topmargin 0pt \oddsidemargin 5mm \renewcommand{%
\thefootnote}{\fnsymbol{footnote}} \newpage \setcounter{page}{0} 
\begin{titlepage}
\begin{flushright}
Berlin Sfb288 Preprint  \\
hep-th/9902011\\
\end{flushright}
\vspace{0.5cm}
\begin{center}
{\Large {\bf The ultraviolet Behaviour of Integrable Quantum Field Theories, 
Affine Toda Field Theory} }

\vspace{1.8cm}
{\large  A. Fring, C. Korff and B.J. Schulz }

\vspace{0.5cm}
{\em Institut f\"ur Theoretische Physik,
Freie Universit\"at Berlin\\ 
Arnimallee 14, D-14195 Berlin, Germany }
\end{center}
\vspace{1.2cm}
 
\renewcommand{\thefootnote}{\arabic{footnote}}
\setcounter{footnote}{0}

\begin{abstract}
We investigate the thermodynamic Bethe ansatz (TBA) equations for a system of 
particles which dynamically interacts via the scattering matrix of affine 
Toda field theory and whose statistical interaction is of a general 
Haldane type. Up to the first leading order, we provide general approximated 
analytical expressions for the solutions of these equations from which we 
derive general formulae for the ultraviolet scaling functions for 
theories in which the underlying Lie algebra is simply laced. 
For several explicit  models we compare the quality of the approximated 
analytical 
solutions against the numerical solutions. We address 
the question of existence and uniqueness of the solutions of the 
TBA-equations, derive 
precise error estimates and  determine the rate of convergence for the applied 
numerical procedure. A general expression for the Fourier transformed 
kernels of the TBA-equations  allows to derive the 
related Y-systems and a reformulation of the equations 
into a universal form.   

\bigskip

\par\noindent
PACS numbers: 11.10Kk, 11.55.Ds, 05.70.Jk, 05.30.-d, 64.60.Fr
\end{abstract}
\vspace{.3cm}
\centerline{February 1999}
\vfill{ \hspace*{-9mm}
\begin{tabular}{l}
\rule{6 cm}{0.05 mm}\\
Fring@physik.fu-berlin.de \\
Korff@physik.fu-berlin.de \\
Schulzb@physik.fu-berlin.de
\end{tabular}}
\end{titlepage}
\newpage 

\section{Introduction}

\noindent More than twenty years ago Yang and Yang \cite{Yang} introduced in
a series of seminal papers a technique which allows to compute thermodynamic
quantities for a system of bosons interacting dynamically via factorizable
scattering. This method was generalized less than ten years ago by Al.B.
Zamolodchikov \cite{TBAZam1} to a system of particles which interact
dynamically in a relativistic manner through a scattering matrix which
belongs to an integrable quantum field theory. This latter approach, usually
referred to as thermodynamic Bethe ansatz (TBA), has triggered numerous
further investigations [3-13]. The reason for this activity is twofold, on
one hand the TBA serves as an interface between conformally invariant
theories and massive integrable deformations of them and on the other hand
it serves as a complementary approach to other methods.

The TBA-approach allows to extract various types of informations from a
massive integrable quantum field theory once its scattering matrix is known.
Most easily one obtains the central charge $c$ of the Virasoro algebra of
the underlying ultraviolet conformal field theory, the conformal dimension $%
\Delta $ and the factor of proportionality of the perturbing operator, the
vacuum expectation values of the energy-momentum tensor $\langle
T_{\,\,\,\mu }^{\mu }\rangle $ and other interesting quantities. Thus, the
TBA provides a test laboratory in which certain conjectured scattering
matrices may be probed for consistency.

In addition, the TBA is useful since it provides quantities which may be
employed in other contexts, like the computation of correlation functions.
For instance the constant of proportionality, the dimension of the
perturbing field and $\langle T_{\,\,\,\mu }^{\mu }\rangle $ may be used in
a perturbative approach around the operator product expansion of a two point
function within a conformal field theory \cite{Zamocorr}. $\langle
T_{\,\,\,\mu }^{\mu }\rangle $ may also be used as an initial value for the
recursive system between different n-particle form factors \cite{Zamocorr}.

In order to obtain data beyond $c$ one has to investigate the behaviour of
the scaling function $c(r)$ ($r$ is the inverse temperature times a mass
scale), which may be viewed as the deformed central charge of the Virasoro
algebra. It will be the central aim of this manuscript to determine this
function. Certain statements in this context can be made in complete
generality, but of course ultimately one has to specify some concrete models
in order to be more explicit. An attractive choice for these models are
affine Toda field theories \cite{ATFT}, since they cover a huge class of
theories due to their Lie algebraic formulation and permit therefore to
extract many universal features. Ultimately they also allow a comparison
against standard perturbation theory in the spirit of \cite{Cassi}.

The first computation of such a scaling function was carried out in \cite
{TBAZam} for the scaling Lee-Yang model and scaling 3-state Potts model
(minimal $A_{2}^{(1)}$-affine Toda field theory). Meanwhile there exist
several computations of this kind \cite{TBAKM,Marcio,Fererati,Saleur}. The general
behaviour one finds is 
\begin{equation}
c(r)=c_{{\rm eff}}+f(r)+\sum_{n=1}^{\infty }c_{n}r^{2n(1-\Delta )}\;\;
\label{typical}
\end{equation}
where $c_{{\rm eff}}=c-24h^{\prime }$ is the effective central charge, with $%
c$ being the usual conformal anomaly, i.e. the central charge of the
Virasoro algebra, and $h^{\prime }$ the lowest conformal dimension of the
underlying conformal field theory. The $c_{n}$ are some constants and
typically $f(r)\sim r^{2}$ or $f(r)\sim r^{2}\ln (r)$. However, in \cite
{ZamoR,MM} a quite different behaviour was observed, namely that $f(r)\sim
(const-\ln (r))^{-2}$, which was attributed therein to zero mode
fluctuations. It is this kind of behaviour which we find for all affine Toda
field theories (see equn. (\ref{universal})). We like to stress that the 
constant $const$ should be different from zero.

In our investigation we slightly modify and generalize an approach which was
introduced originally in \cite{ZamoR}. We also include in our analysis the
possibility that the statistical interaction is of general Haldane type \cite
{Haldane} and investigate the TBA-equations adapted to this situation \cite
{BF}. We investigate in detail the expressions for the TBA-kernels and their
Fourier transformed versions, from which we obtain a universal form for the
relevant TBA-equations. We derive the related Y-systems. Furthermore, we
address the question of the existence and uniqueness of the solutions of the
TBA equations, derive error estimates and the rate of convergence for the
numerical procedure applied. Our analytical considerations culminate with
the derivation of a general expression for the scaling function valid for
all affine Toda field theories in which the underlying Lie algebra is simply
laced (\ref{universal}) and which depends in a universal manner only on the
rank of the algebra, its Coxeter number and the effective coupling constant.
We compare the general formulae against the numerical solutions for several
explicit models, i.e. for the $%
A_{1}^{(1)},A_{2}^{(1)},A_{2}^{(2)},A_{3}^{(1)}$ and $%
(G_{2}^{(1)},D_{4}^{(3)})$-affine Toda field theories.

Our manuscript is organized as follows: In section 2 we recall the
TBA-equation for general Haldane statistics and explain how it can be used
to compute explicitly expressions for scaling functions. In section 3 we
derive general expression for the approximated analytical expression for the
solutions of the TBA-equations and for the scaling functions $c(r)$. In
section 4 we concretely test our statements for a system involving the
scattering matrix of affine Toda field theories. We compare the general
approximated analytical expression with the numerical solution for some
explicit models. In section 5 we address the question of the existence and
uniqueness of the solutions of the TBA equations and derive error estimates
and the rate of convergence. We state our conclusions in section 6.

\section{The TBA-equations}

\noindent We consider a multi-particle system involving $l$ different types
of particles and assume that the two-particle scattering matrix $%
S_{ij}(\theta )$ (as a function of the rapidity $\theta $) together with the
corresponding mass spectrum ($m_{i}$ is the mass of particle type $i$) have
been determined. The $l$ coupled TBA-equations which describe such a
system, based on the assumption that the dynamical interaction is
characterized by the scattering matrices $S_{ij}(\theta )$ and whose
statistical interaction is governed by general Haldane statistics \cite
{Haldane} $g_{ij}$\footnote{%
The bosonic and fermionic statistics correspond to $g_{ij}=0$ and $%
g_{ij}=\delta _{ij}$, respectively.}, read \cite{BF} 
\begin{equation}
rm_{i}\cosh \theta +\ln \left( 1-e^{-L_{i}(\theta ,r)}\right)
=\sum\limits_{j=1}^{l}\left( \Phi _{ij}*L_{j}\right) (\theta ,r)\quad .
\label{HTBA}
\end{equation}
The scaling parameter $r$ is  the inverse temperature times a mass scale and
we denote as usual the convolution of two functions $f$ and $g$ by $%
(f*g)(\theta )$ $:=1/(2\pi )\int d\theta ^{\prime }f(\theta -\theta ^{\prime
})g(\theta ^{\prime })$ and the kernel by 
\begin{equation}
\Phi _{ij}(\theta ):=-i\frac{d}{d\theta }\ln S_{ij}(\theta )-2\pi
g_{ij}\delta \left( \theta \right) =\varphi _{ij}(\theta )-2\pi g_{ij}\delta
\left( \theta \right) \quad .  \label{inva}
\end{equation}
We recall the important fact that the functions $L_{i}(\theta ,r)$ are
related to the ratio of densities $\rho _{r}^i$of particles inside the
system over the densities of $\rho _{h}^i$ of available states as $%
L_{i}(\theta ,r)$ $=\ln (1+\rho_{r}^i/\rho_{h}^i)$. This
implies that $L_{i}(\theta ,r)\geq 0$, a property we will frequently appeal
to. In addition we will make use of the fact that $L_i(\theta ,r)$ is an even
function in $\theta $. Once the system of coupled non-linear integral
equations (\ref{HTBA}) is solved (usually this may only be achieved
numerically) for the $l$ unknown functions $L_{i}(\theta ,r)$, one is in a
position to determine the scaling function 
\begin{equation}
c(r)=\frac{6r}{\pi ^{2}}\sum\limits_{i=1}^{l}m_{i}\int\limits_{0}^{\infty
}d\theta L_{i}(\theta ,r)\cosh \theta \quad ,  \label{c(r)}
\end{equation}
which may be viewed as the off-critical central charge of the Virasoro
algebra. We shall now adopt the method of \cite{ZamoR} and instead of
regarding the TBA-equation as an integral equation, we transform it into an
infinite order differential equation. This is possible with the sole
assumption that the Fourier transform of the function $\Phi _{ij}(\theta )$
can be expanded as a power series 
\begin{equation}
\widetilde{\Phi }_{ij}(k)=\int\limits_{-\infty }^{\infty }d\theta \Phi
_{ij}(\theta )e^{ik\theta }=2\pi \sum_{n=0}^{\infty }(-i)^{n}\eta
_{ij}^{(n)}k^{n}\quad .  \label{Fexp}
\end{equation}
Then it is a simple consequence of the convolution theorem\footnote{$%
(f*g)(\theta )$ $=1/(2\pi )^{2}\int dk\widetilde{f}(k)\widetilde{g}%
(k)e^{-ik\theta }$} that the integral equation (\ref{HTBA}) may also be
written as an infinite order differential equation 
\begin{equation}
rm_{i}\cosh \theta +\ln \left( 1-e^{-L_{i}(\theta ,r)}\right)
=\sum\limits_{j=1}^{l}\sum\limits_{n=0}^{\infty }\eta
_{ij}^{(n)}L_{j}^{(n)}(\theta ,r)\quad .  \label{TBADiff}
\end{equation}
We introduced here the abbreviation $L_i^{(n)}(\theta ,r)=(d/d\theta
)^{n}L_i(\theta ,r)$. The two alternative formulations of the TBA-equations
serve different purposes. Its variant in form of an integral equation (\ref
{HTBA}) is most convenient for numerical studies, whereas the differential
equations (\ref{TBADiff}) turn out to be useful for analytical
considerations.

From a numerical point of view it is advantageous to make a change of
variables and instead of the functions $L_{i}(\theta ,r)$ use the so-called
pseudo-energies $\epsilon _{i}(\theta ,r)$ defined via the equations 
\begin{equation}
\epsilon _{i}(\theta ,r):=-\sum\limits_{j}g_{ij}L_{j}(\theta ,r)-\ln \left(
1-e^{-L_{i}(\theta ,r)}\right) \;.  \label{pseudoe}
\end{equation}
For fermionic and bosonic type of statistics the formulation of the
TBA-equations in terms of pseudo-energies is most common. However, since (%
\ref{pseudoe}) may obviously not be solved in general, it is more convenient
to keep the TBA-equations in terms of the $L_{i}(\theta ,r)$ for general
Haldane type of statistics. Alternatively we may relate all types of statistics
to the fermionic one in the following way. Considering a system which
interacts statistically via $g_{ij}$, we can define the quantity $%
g_{ij}^{\prime }=\delta _{ij}-g_{ij}$ and parameterize all $L-$functions by $%
L_{i}(\theta ,r)=:{\cal L}_{i}(\theta ,r)=\ln \left( 1+e^{-\varepsilon
_{i}(\theta ,r)}\right) $. Then the TBA-equations (\ref{HTBA}) may be
rewritten as 
\begin{equation}
\varepsilon _{i}(\theta ,r)=rm_{i}\cosh \theta -\sum\limits_{j=1}^{l}\left[
\left( (\varphi _{ij}+2\pi g_{ij}^{\prime }\delta )*{\cal L}_{j}\right)
(\theta ,r)\right] \,.  \label{eps}
\end{equation}
It should be kept in mind that $\varepsilon _{i}(\theta ,r)$ ($\neq \epsilon
_{i}(\theta ,r)!$) is now a formal parameter and, except in the fermionic
case, it is not related to the ratio of particle densities 
$\rho_{r}^i/\rho_{h}^i$ in the characteristic way as a distribution function
associated to the relevant statistics, i.e. $\epsilon _{i}(\theta ,r)$
obtained as a solution of the equations (\ref{pseudoe}).

\section{The ultraviolet Limit}

\noindent We shall now analytically investigate the behaviour of the scaling
function in the ultraviolet limit, i.e. $r$ is going to zero. For this
purpose we generalize the procedure of Zamolodchikov \cite{ZamoR}, which
leads to approximated analytical expressions. We attempt to keep the
discussion model independent and free of a particular choice of the
statistical interaction as long as possible. Introducing the quantity 
$\hat{L}_i(\theta ):=L_i(\theta -\ln (r/2),r)$ and 
performing the shift $\theta
\rightarrow \theta -\ln (r/2),$ the TBA-equations (\ref{TBADiff}) acquire
the form 
\begin{equation}
m_{i}e^{\theta }+\ln \left( 1-e^{-\hat{L}_{i}(\theta )}\right)
=\sum\limits_{j=1}^{l}\sum\limits_{n=0}^{\infty }\eta _{ij}^{(n)}\hat{L}%
_{j}^{(n)}(\theta )\,\,.  \label{TBAshift}
\end{equation}
Here we have neglected the terms proportional to $e^{2\ln (r/2)-\theta }$,
under the assumption that $2\ln (r/2)\ll \theta $. Obviously the $r$%
-dependence has vanished, such that the $\hat{L}_{i}(\theta )$ are $r$%
-independent. The equation for the scaling function (\ref{c(r)}) becomes,
under similar manipulations and the neglect of similar terms as in the
derivation of (\ref{TBAshift}), 
\begin{equation}
c(r)=\frac{6}{\pi ^{2}}\sum\limits_{i=1}^{l}m_{i}\int\limits_{\ln
(r/2)}^{\infty }d\theta \hat{L}_{i}(\theta )e^{\theta }\quad .
\label{cshift}
\end{equation}

In analogy to the procedure of \cite{ZamoR}, we consider now the so-called
``truncated scaling function'' 
\begin{equation}
\hat{c}(r,r^{\prime })=\frac{6}{\pi ^{2}}\sum\limits_{i=1}^{l}m_{i}\int%
\limits_{r^{\prime }}^{\infty }d\theta \hat{L}_{i}(\theta )e^{\theta }\quad ,
\label{truncc}
\end{equation}
which obviously coincides with $c(r)$ for $r^{\prime }=$ $\ln (r/2)$. At
this point we make several assumptions:

\begin{enumerate}
\item[i)]  The functions $\hat{L}_{i}(\theta )$ obey the equations (\ref
{TBAshift}).

\item[ii)]  In the power series expansion (\ref{Fexp}) the coefficients are
symmetric in the particle type indices, i.e. $\eta _{ij}^{(n)}=$ $\eta
_{ji}^{(n)}$.

\item[iii)]  All odd coefficients vanish in (\ref{Fexp}), i.e. $\eta
_{ij}^{(2n+1)}=0$.

\item[iv)]  The asymptotic behaviour of the function $\hat{L}_{i}(\theta )$
and its derivatives read 
\begin{eqnarray}
\lim_{\theta \rightarrow \infty }\hat{L}_{i}^{(n)}(\theta ) &=&0\qquad \text{%
for }n\geq 1\,,1\leq i\leq l\quad ,  \label{a1} \\
\lim_{\theta \rightarrow \infty }e^{\theta }\hat{L}_{i}(\theta ) &=&0\qquad 
\text{for }1\leq i\leq l\quad .  \label{a2}
\end{eqnarray}
\end{enumerate}

Assumption ii) is guaranteed when the two-particle scattering matrix is
parity invariant and the statistical interaction is symmetric in particle
type, $g_{ij}=g_{ji}$. The requirement iii) puts of course constraints on the
scattering matrices for given statistics, or vice versa. It will turn out in
the next section, that for fermionic statistics it is satisfied by all
scattering matrices of interest to us. The asymptotic behaviour iv) will be
verified in retrospective, that is all known numerical solutions exhibit
this kind of asymptotic behaviour.

Under these assumptions the truncated scaling function may also be written
as 
\begin{eqnarray}
\hat{c}(r,r^{\prime }) &=&\frac{3}{\pi ^{2}}\sum\limits_{i,j=1}^{l}\left(
\sum\limits_{n=1}^{\infty }\eta
_{ij}^{(2n)}\sum\limits_{k=1}^{2n-1}(-1)^{k+1}\hat{L}_{i}^{(k)}(r^{\prime })%
\hat{L}_{j}^{(2n-k)}(r^{\prime })+\eta _{ij}^{(0)}\hat{L}_{i}(r^{\prime })%
\hat{L}_{j}(r^{\prime })\right)  \nonumber \\
&&\!\!\!-\frac{6}{\pi ^{2}}\sum\limits_{i=1}^{l}\left( L(1-e^{-\hat{L}%
_{i}(r^{\prime })})+\frac{\hat{L}_{i}(r^{\prime })}{2}\ln (1-e^{-\hat{L}%
_{i}(r^{\prime })})+m_{i}e^{r^{\prime }}\hat{L}_{i}(r^{\prime })\right) .
\label{Solu}
\end{eqnarray}
Here $L(x)=\sum_{n=1}^{\infty }x^{n}/n^{2}+1/2\ln (x)\ln (1-x)$ denotes
Rogers dilogarithm (e.g. \cite{Lewin}). The equality (\ref{Solu}) is most
easily derived when considering the differential equation which results from
(\ref{truncc}) 
\begin{equation}
\frac{\partial \hat{c}(r,r^{\prime })}{\partial r^{\prime }}=-\frac{6}{\pi
^{2}}e^{r^{\prime }}\sum\limits_{i=1}^{l}m_{i}\hat{L}_{i}(r^{\prime })\quad .
\label{Diff}
\end{equation}
One may now verify by direct substitution\footnote{%
The identity $\int\limits_{0}^{x}dt\ln (1-e^{-t})=L(1-e^{-x})+x/2\ln
(1-e^{-x})$ is useful in this context.} that (\ref{Diff}) is solved by (\ref
{Solu}) when i)-iii) hold. The constant of integration is fixed by the
property $\hat{c}(r,\infty )=0$, such that (\ref{Solu}) is exact if iv)
holds.

\subsection{The extreme Limit}

One of the best known outcomes of the TBA-analysis is the fact that in the
extreme ultraviolet limit the scaling function becomes the effective central
charge of the underlying ultraviolet conformal field theory, i.e. $%
\lim_{r\rightarrow 0}c(r)=c_{\text{eff}}$. In order to carry out this limit,
we note that the assumptions (\ref{a1}) and (\ref{a2}) also imply 
\begin{equation}
\lim_{r,\theta \rightarrow 0}L_{i}^{(n)}(\theta ,r)=0\qquad \text{for }n\geq
1\,,1\leq i\leq l\quad ,  \label{slowvary}
\end{equation}
such that equations (\ref{TBADiff}) become a set of coupled non-linear
equations for the $l$ constants $L_{i}(0,0)$%
\begin{equation}
\ln \left( 1-e^{-L_{i}(0,0)}\right) =\sum\limits_{j=1}^{l}\eta
_{ij}^{(0)}L_{j}(0,0)\quad .  \label{ex1}
\end{equation}
Due to the fact that $L_{i}(\theta ,r)$ is positive, we deduce that these
equations admit physical solutions when $\eta _{ij}^{(0)}\leq 0$. For a
given S-matrix this condition will restrict possible statistical
interactions. With the help of the solutions of (\ref{ex1}) we recover from (%
\ref{Solu}) the well known formula for the extreme ultraviolet limit 
\begin{equation}
\lim\limits_{r\rightarrow 0}c(r)=c_{\text{eff}}=\frac{6}{\pi ^{2}}%
\sum\limits_{i=1}^{l}L(1-e^{-L_{i}(0,0)})\quad .  \label{cex}
\end{equation}
These equations have been analyzed extensively in the literature for various
models \cite{TBAZam,TBAKM,DoRa,TBA} and as we demonstrated, also hold 
for general (Haldane) type of statistics.

\subsection{Next leading Order}

\noindent In order to keep the next leading order in the ultraviolet limit
we proceed as follows: Under the assumption that the symmetry property of 
$\hat{L}_i(\theta )$ still holds at this point of the derivation, we may
neglect in (\ref{TBAshift}) also the term $e^{\theta }$. Assuming that $\hat{%
L}_{i}(\theta )$ is large and (\ref{a1}) holds, the TBA-equation in the
ultraviolet limit may be approximated by 
\begin{equation}
\sum\limits_{j=1}^{l}(\eta _{ij}^{(2)}\hat{L}_{j}^{(2)}(\theta )+\eta
_{ij}^{(0)}\hat{L}_{j}(\theta ))+e^{-\hat{L}_{i}(\theta )}=0\quad .
\label{APPdiff}
\end{equation}
Unfortunately this equation may not be solved analytically in its full
generality. However, choosing now the statistics in such a way that the $%
\eta _{i}^{(0)}=\sum_{j=1}^{l}\eta _{ij}^{(0)}=0$ we may solve (\ref{APPdiff}%
). In most cases we shall be considering in the following, this condition
implies that we have fermionic type of statistics. For this case the
solution of (\ref{APPdiff}) reads 
\begin{equation}
\hat{L}_{i}(\theta )=\ln \left( \frac{\sin ^{2}\left( \alpha _{i}\left(
\theta -\beta _{i}\right) \right) }{2\alpha _{i}^{2}\eta _{i}^{(2)}}\right)
+\ln \left( \frac{\cos ^{2}\left( \widetilde{\alpha }_{i}\left( \theta -%
\widetilde{\beta }_{i}\right) \right) }{2\widetilde{\alpha }_{i}^{2}\eta
_{i}^{(2)}}\right) \quad ,  \label{Soll}
\end{equation}
with $\alpha _{i},\beta _{i},\widetilde{\alpha }_{i},\widetilde{\beta }_{i}$
being the constants of integration and $\eta _{i}^{(2)}=\sum_{j=1}^{l}\eta
_{ij}^{(2)}$. Note that the assumption $\eta _{i}^{(2)}\neq 0$ is not always
guaranteed below. We will discard the second term in the following w.l.g.
Under the same assumptions the expression for the truncated scaling function
(\ref{Solu}) becomes 
\begin{equation}
\hat{c}(r,r^{\prime })=l+\frac{3}{\pi ^{2}}\sum\limits_{i=1}^{l}\left( \eta
_{i}^{(2)}\left( \hat{L}_{i}^{(1)}(r^{\prime })\right) ^{2}-2e^{-\hat{L}%
_{i}(r^{\prime })}\right) \quad .  \label{ctrunc}
\end{equation}
Substitution of the solution (\ref{Soll}) into (\ref{ctrunc}) yields 
\begin{equation}
\hat{c}(r,r^{\prime })=l-\frac{12\,}{\pi ^{2}}\sum\limits_{i=1}^{l}\eta
_{i}^{(2)}\alpha _{i}^{2}\quad .  \label{ctrun1}
\end{equation}
Notice that this expression for the truncated effective central charge is
independent of the constants $\beta _{i}$ and $r^{\prime }$. We use the
latter property to argue that in fact the r.h.s. of (\ref{ctrun1})
corresponds to the scaling function $c(r)$. Invoking now also the property $%
\hat{L}_{i}(\theta )=\hat{L}_{i}(2\ln (r/2)-\theta )$ we obtain the
additional relations 
\begin{equation}
\alpha _{i}=\frac{n\pi }{2(\beta _{i}-\ln (r/2))}\quad ,
\end{equation}
with $n$ being an odd integer, which we choose to be one. At the moment we
do not have any further argument at hand in order to fix the remaining
constant. Therefore we have 
\begin{eqnarray}
L_{i}(\theta ,r) &=&\ln \left( \frac{\cos ^{2}\left( \alpha _{i}\theta
\right) }{2\alpha _{i}^{2}\eta _{i}^{(2)}}\right)  \label{Lappp} \\
L_{i}^{(1)}(\theta ,r) &=&-2\alpha _{i}\tan (\alpha _{i}\theta )\quad \quad
\label{L1} \\
L_{i}^{(2)}(\theta ,r) &=&-2\alpha _{i}^{2}/\cos ^{2}(\alpha _{i}\theta ) \\
L_{i}^{(3)}(\theta ,r) &=&-4\alpha _{i}^{3}\tan (\alpha _{i}\theta )/\cos
^{2}(\alpha _{i}\theta ) \\
L_{i}^{(n)}(\theta ,r) &\sim &\alpha _{i}^{n}\quad .  \label{L4}
\end{eqnarray}
Using the fact that $\alpha _{i}$ tends to zero for small $r$, the equations
(\ref{L1})-(\ref{L4}) demonstrate the consistency with the assumption that
the derivatives of $L_{i}(\theta ,r)$ with respect to $\theta $ are
negligible, i.e. equation (\ref{slowvary}). Closer inspection shows that for
given $r$ the series build from the $L_{i}^{(n)}(\theta ,r)$ starts to
diverge at a certain value of $n$. Since (\ref{Lappp}) is not exact this
does not pose any problem, but one should be aware of it. The scaling
function becomes in this approximation 
\begin{equation}
c(r)=l-\,3\,\,\sum\limits_{i=1}^{l}\frac{\eta _{i}^{(2)}}{(\beta _{i}-\ln
(r/2))^{2}}\quad .  \label{cappp}
\end{equation}
As already pointed out there does not seem to be any argument in this
approach which allows to fix the constant $\beta _{i}$. However, we will
present below a natural guess and also resort to numerical data to fix it.

In order to perform more concrete computations one has to specify a
particular model at this point.

\section{Affine Toda Field Theory}

Affine Toda field theories \cite{ATFT} constitute a huge, important and well
studied class of relativistically invariant integrable models in 1+1
dimensions. The exact two-particle scattering matrices of all affine Toda
field theories with real coupling constant were constructed on the base of
the bootstrap principle \cite{TodaS} and its generalized version \cite{nons}%
. Various cases have been checked perturbatively. The theories exhibit an
entirely different behaviour depending on whether they are self-dual or not.
Here duality has a double meaning, on one hand it refers to the invariance
of the algebra under the interchange of roots and co-roots (i.e. $%
A_{n}^{(1)},A_{2n}^{(2)},D_{n}^{(1)},E_{6}^{(1)},E_{7}^{(1)},E_{8}^{(1)}$)
and on the other hand it refers to the strong-weak duality in the coupling
constant.

\subsection{The S-matrix}

\subsubsection{Simply laced Lie algebras}

We recall now some well known facts about the two particle scattering matrix
and present some new features which will be important for our analysis
below. For theories related to simply laced Lie algebras the two particle
S-matrices \cite{TodaS,PD,FO} involving the $r$  
 different types of particles may be furnished into the universal
form 
\begin{equation}
S_{ij}\left( \theta \right) =\prod\limits_{q=1}^{h}\left\{ 2q-\frac{c(i)+c(j)%
}{2}\right\} _{\theta }^{-\frac{1}{2}\lambda _{i}\cdot \sigma ^{q}\gamma
_{j}},\qquad 1\leq i,j\leq r={\rm rank }{\bf g}  \,.  \label{S}
\end{equation}
The building blocks $\left\{ x\right\} _{\theta }$ may be expressed in terms
of hyperbolic functions, albeit it will be most convenient for our purposes
to use their integral representation 
\begin{equation}
\left\{ x\right\} _{\theta }=\exp \int\limits_{0}^{\infty }\frac{dt}{t\sinh t%
}\,\,f_{x,B}(t)\sinh \left( \frac{\theta t}{i\pi }\right)  \label{block}
\end{equation}
with 
\begin{equation}
\,f_{x,B}(t)=8\sinh \left( \frac{tB}{2h}\right) \sinh \left( \frac{t}{2h}%
\left( 2-B\right) \right) \sinh \left( \frac{t}{h}\left( h-x\right) \right)
\,\,.\quad
\end{equation}
We adopt here the notations of \cite{FO} and denote the Coxeter number by $h$%
, fundamental weights by $\lambda _{i}$, the colour values related to the
bicolouration of the Dynkin diagram by $c(i)$ and a simple root $\alpha _{i}$
times $c(i)$ by $\gamma _{i}$. The Coxeter element is chosen to be $\sigma
=\sigma _{-}\sigma _{+}$, where the elements $\sigma _{\pm }=\prod_{i\in
\Delta _{\pm }}\sigma _{i}$ are introduced, with $\sigma _{i}$ being a Weyl
reflections and $\Delta _{\pm }$ the set of simple roots with colour values $%
c(i)=\pm 1$, respectively. The effective coupling constant $B(\beta )=(\beta
^{2}/2\pi )/(1+\beta ^{2}/4\pi )$ depends monotonically on the coupling
constant $\beta $ and takes values between 0 and 2.

Many of the properties of (\ref{S}) are very well documented in the
literature \cite{TodaS,PD,FO} and we will therefore only concentrate on
those which are relevant for our investigations. One of the remarkable
features is the strong-weak duality in the coupling constant $\beta $, which
is a particular example for one of the dualities which are currently
ubiquitous in string theoretical investigations. Whilst in the latter
context duality serves as a powerful principle, we will frequently employ it
as a simple consistency check on various expressions which arise during our
computations. The strong-weak duality manifests itself by the fact that each
individual building block, and consequently the whole S-matrix, possesses
the symmetry $B\rightarrow 2-B$. A further check, which we wish to pursue from
time to time for consistency reasons, is taking the coupling constant to
zero, i.e. $B(0)=0$ or equivalently $B(\infty )=2$, such that we obtain a
free theory with $S_{ij}(\theta )=1$. Also we want to compare with some
known results for the so-called minimal part\footnote{
This part satisfies by itself the bootstrap equations and the additional
factor is of a CDD-nature, meaning that is does not produce any relevant
poles inside the physical sheet.} of the scattering matrix (\ref{S}), which
is formally obtained by taking the limit $B\rightarrow i\infty \footnote{
For the blocks $\left\{ x\right\} _{\theta }$ in form of hyperbolic
functions this limit is trivial. For the integral representation (\ref{block}
) it requires a bit more effort, but is easily verified with the help of the
Riemann-Lebesgue theorem (If $g(x)\in L_{1}(-\infty ,\infty )$ then $
\lim\limits_{t\rightarrow \pm \infty }\int_{-\infty }^{\infty
}g(x)e^{-itx}dx=0$).}.$

It will be important below to recall that the block $\left\{ 1\right\}
_{\theta }$ only occurs (with power one) in the scattering matrix between
two particles of the same type, i.e. in $S_{ii}(\theta )$, which implies for
the S-matrices (\ref{S}) $S_{ij}(0)=(-1)^{\delta _{ij}}$. A further relation
which exhibits the occurrence of particular blocks is 
\begin{equation}
S_{ij}\left( \theta \right) =\left\{ 2\right\} _{\theta
}^{I_{ij}}\prod\limits_{q=2}^{h-2}\left\{ 2q-\frac{c(i)+c(j)}{2}\right\}
_{\theta }^{-\frac{1}{2}\lambda _{i}\cdot \sigma ^{q}\gamma _{j}},\qquad
c(i)\neq c(j)\text{\thinspace .}  \label{2block}
\end{equation}
Here $I_{ij}$ is the incidence matrix of the Lie algebraic Dynkin diagram,
i.e. twice the unit matrix minus the Cartan matrix $K$. Since relation (\ref
{2block}) will be important for our analysis and does not seem to appear in
the literature, we will briefly prove it here. We make use of the
interrelation between the incidence matrix and particular elements of the
Weyl group and the action of these elements on simple roots \cite{MOO,FO} 
\begin{equation}
\sum\limits_{j=1}^{r}I_{ij}\lambda _{j}=(\sigma _{-}+\sigma _{+})\lambda
_{i}\quad \text{and \quad }\sigma _{c(i)}\alpha _{j}=-\sigma
^{(c(j)-c(i))/2}\alpha _{j}\,\,.  \label{hh}
\end{equation}
Taking the inner product of the first equation in (\ref{hh}) with $\alpha
_{j}$ and the subsequent application of the second relation on the r.h.s.
yields 
\begin{equation}
I_{ij}=\lambda _{i}\cdot \sigma _{-c(j)}\alpha _{j}+\lambda _{i}\cdot \sigma
_{c(j)}\alpha _{j}=-\lambda _{i}\cdot \alpha _{j}-\lambda _{i}\cdot \sigma
^{c(j)}\alpha _{j}\,\,.
\end{equation}
Therefore, by the orthogonality of fundamental weights and simple roots and
the symmetry in $i$ and $j$, it follows directly that 
\begin{equation}
\lambda _{i}\cdot \sigma ^{\pm 1}\gamma _{j}=\mp I_{ij}\,,\qquad c(i)\neq
c(j)\quad .
\end{equation}
Noting further that $\left\{ x\right\} _{\theta }=\left\{ 2h-x\right\}
_{\theta }^{-1}$, we may extract the block $\left\{ 2\right\} _{\theta }$
with power $I_{ij}$ from the product in (\ref{S}) and obtain (\ref{2block}).

Remarkably one may also carry out the product in (\ref{S}) 
and obtain a closed expression for $S$ in the form
\begin{equation}
S_{ij}\left( \theta \right) =\exp \left( i\int\limits_{0}^{\infty }\frac{dt%
}{t}\left( f_{h+h\pi/(2t)),B}(t)   
\left( 2\cosh \frac{\pi t}{h}-I\right) _{ij}^{-1}-2\delta
_{ij}\right) \sin \left( \frac{\theta t}{\pi }\right) \right) \,.  \label{SC}
\end{equation}
The Lie algebraic quantities  involved in this identity, i.e. $I$ and $h$,
are more easily accessible than the orbits of the Coxeter elements,
however, this is at the cost that the singularity structure is less
transparent.  Formula (\ref{SC}) coincides  with equation (5.2) in \cite
{Oota} (up to a factor $(-1)^{\delta _{ij}}$), once the general expression
in there is reduced to the simply laced case.

In the course of our argumentation we will also employ the identity 
\begin{equation}
S_{ij}\left( \theta -\frac{i\pi }{h}\right) S_{ij}\left( \theta +\frac{i\pi 
}{h}\right) =\prod\limits_{l=1}^{r}S_{il}\left( \theta \right)
^{I_{lj}}\qquad \theta \neq 0\,\,,  \label{rel}
\end{equation}
which was derived first in \cite{Rava}.

\subsubsection{Non-simply laced Lie algebras}

Once one turns to the consideration of affine Toda field theories related to
non-simply laced Lie algebras many of the universal features which one
observes for simply laced Lie algebras cease to be valid. For instance for
the simply laced Lie algebras it is known that the singularities of (\ref{S}%
) inside the physical sheet do not depend on the coupling constant and
remarkably all masses renormalise with an overall factor. These latter
behaviours dramatically change once the related Lie algebras are taken to be
non-simply laced (except $A_{2n}^{(2)}$ which is also self-dual). In fact in
these cases one can not associate anymore a unique Lie algebra to the
quantum field theory, but a dual pair related to each other by the
interchange of roots and co-roots. General formulae similar to (\ref{S}),
which are valid for all non-simply Lie laced algebras have not been
constructed yet. However, one can still construct S-matrices 
\cite{nons,nons2} 
\begin{equation}
S_{ij}\left( \theta \right) =\prod\limits_{x,y}\left[ x,y\right] _{\theta }
\label{nichtsimp}
\end{equation}
out of some universal building blocks which generalize (\ref{block}) 
\begin{equation}
\left[ x,y\right] _{\theta }=\exp \int\limits_{0}^{\infty }\frac{dt}{t\sinh t%
}\,\,g_{x,y,B}(t)\sinh \left( \frac{\theta t}{i\pi }\right) \,\,.
\label{nonblocks}
\end{equation}
In distinction to the simply laced case, the variable $x$ may now become
non-integer, $y$ is $0,1/2,1$ or $-1/4$ and 
\begin{equation}
g_{x,y,B}(t)=8\sinh \left( \frac{tB}{2H}(1-2y)\right) \sinh \left( \frac{t}{%
2H}\left( 2-B\right) \right) \sinh \left( \frac{t}{H}\left( H-x\right)
\right) \quad .
\end{equation}
The Coxeter number $h$
has been replaced by a ``floating Coxeter number '' $H$, in the sense that
it equals the Coxeter number of one of the two dual algebras in the weak
limit and the other in the strong limit. The singularities of (\ref
{nichtsimp}) now depend on the coupling constant which leads to a
modification of the bootstrap as explained in \cite{nons}. The
shift of the masses resulting from the renormalization procedure
can not be compensated anymore by an overall factor. 

In contrast to the simply laced case, the control over the set in which 
$x$ and $y$ take their values is slightly different than in (\ref{S}).
Whereas in (\ref{S}) the powers of the blocks may be computed directly
from the Lie algebraic quantities, in the non-simply laced case they are
obtained solving at first some recursive equations  and 
thereafter extracting the powers from a generating function 
\cite{Oota}. There exists however a 
generalization of (\ref{SC}) which also includes the non-simply laced case.  
In this paper we will not treat the completely generic case and shall be
content with treatment of one particular 
case in order to exhibit the difference
towards the simply laced case, such that (\ref{nichtsimp}) and 
(\ref{nonblocks}) are sufficient for our purposes.

\subsection{The TBA-kernels}

Given the S-matrices we are now in a position to compute the relevant
quantities for our analysis, i.e. the kernels appearing in the TBA-equations
(\ref{HTBA}) and the Fourier coefficients in equation (\ref{Fexp}).

\subsubsection{Simply laced Lie algebras}

Taking the logarithmic derivative of the phase of the scattering matrix (\ref
{S}) the TBA-kernels (without the statistics factor) are easily computed to 
\begin{equation}
\varphi _{ij}(\theta )=-\frac{1}{2}\sum\limits_{q=1}^{h}(\lambda _{i}\cdot
\sigma ^{q}\gamma _{j})\omega _{2q-(c(i)+c(j))/2}(\theta )\quad ,
\label{kernel}
\end{equation}
where 
\begin{eqnarray}
\omega _{x}(\theta ) &=&-i\frac{d}{d\theta }\ln \left\{ x\right\} _{\theta }=%
\frac{1}{\pi }\int\limits_{0}^{\infty }\frac{dt}{\sinh t}\,\,f_{x,B}(t)\cos
\left( \frac{\theta t}{\pi }\right) ,  \label{intrep} \\
&=&\gamma _{x-1}(\theta )+\gamma _{x+1}(\theta )-\gamma _{x+B-1}(\theta
)-\gamma _{x-B+1}(\theta )\,.  \label{hyprep}
\end{eqnarray}
We introduced the function $\gamma _{x}(\theta )=\sin (\pi x/h)/(\cos (\pi
x/h)-\cosh \theta )$ and set $\gamma _{0}(\theta )=0$. Depending on the
context either the variant (\ref{intrep}) or (\ref{hyprep}) turn out to be
more convenient. The Fourier transform is most easily evaluated when we use
the integral representation (\ref{intrep}) for each block 
\begin{eqnarray}
\widetilde{\omega }_{x}(k) &=&-\frac{\pi \,f_{x,B}(\pi k)}{\sinh \pi k}\,
\label{genexp} \\
&=&\widetilde{\gamma }_{x-1}(k)+\widetilde{\gamma }_{x+1}(k)-\widetilde{%
\gamma }_{x+B-1}(k)-\widetilde{\gamma }_{x-B+1}(k),
\end{eqnarray}
where $\widetilde{\gamma }_{x}(k)=2\pi \sinh [(x/h-1)\pi k]/\sinh (\pi k)$.
Therefore we have 
\begin{equation}
\widetilde{\varphi }_{ij}(k)=-\frac{1}{2}\sum\limits_{q=1}^{h}(\lambda
_{i}\cdot \sigma ^{q+(c(j)-1)/2}\gamma _{j})\widetilde{\omega }%
_{2q+(c(j)-c(i))/2-1}(k)\,\,.  \label{FTS}
\end{equation}
This means, that just like the S-matrix itself, we may express the
TBA-kernels and their Fourier transformed versions in terms of general
blocks. The difference is that now instead of the powers of the blocks their
pre-factors are determined by orbits of the Coxeter element. Notice also
that the property $\left\{ x\right\} _{\theta }=\left\{ 2h+x\right\}
_{\theta }$ guaranteed that in (\ref{S}) the products $\prod_{q=1}^{h}$ and $%
\prod_{q=0}^{h-1}$ are equivalent. Since now $\widetilde{\gamma }_{x}(k)\neq 
\widetilde{\gamma }_{2h+x}(k)$, we have to take the sum over the appropriate
range. This is the reason why $q$ is shifted in (\ref{FTS}). Remarkably for
the Fourier transformed kernels it is possible to evaluate the whole sum
over $q$ in (\ref{FTS}) in an indirect way. We obtain the universal form 
\begin{eqnarray}
\widetilde{\varphi }_{ij}(k) &=&2\pi \sum\limits_{l=1}^{r}\left( I-2\cosh 
\frac{\pi }{h}k\right) _{il}^{-1}\left( I-2\cosh \frac{\pi }{h}k(1-B)\right)
_{lj}\,\,  \label{funi} \\
&=&8\pi \sinh \left( \frac{(B-2)\pi k}{2h}\right) \sinh \left( \frac{B\pi k}{%
2h}\right) \left( 2\cosh \frac{\pi k}{h}-I\right) _{ij}^{-1}-2\pi \delta
_{ij} \,\, .  \label{funii}
\end{eqnarray}

A similar matrix identity has turned out to be extremely useful in the
investigation \cite{TBAZamun} of the TBA-kernel for the minimal part of the
scattering matrix (\ref{S}). Equation (\ref{funi}) coincides with an
identity quoted in \cite{MM} once the roaming parameter therein is chosen
in such the way that the resonance scattering matrices  take on
the form of (\ref{S}). Taking the inverse Fourier transform of (\ref{funii})
we obtain after an intgeration with respect to $\theta$ the integral 
representation (\ref{SC}) for the scattering matrix.

\begin{figure}[h]
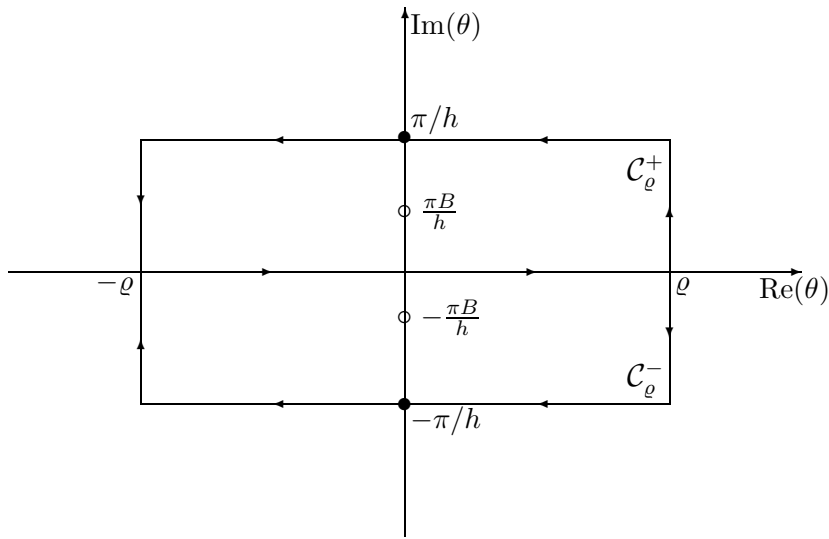

\[
\put(-150,0){\vector(1,0){300}} \put(0,-100){\vector(0,1){200}}
\put(-100,-50){\line(0,1){100}} \put(100,-50){\line(0,1){100}}
\put(-100,50){\line(1,0){200}} \put(-100,-50){\line(1,0){200}}
\put(100,0){\vector(0,1){25}} \put(100,0){\vector(0,-1){25}}
\put(-100,50){\vector(0,-1){25}} \put(-100,-50){\vector(0,1){25}}
\put(0,50){\vector(-1,0){50}} \put(0,-50){\vector(-1,0){50}}
\put(100,50){\vector(-1,0){50}} \put(100,-50){\vector(-1,0){50}}
\put(0,0){\vector(1,0){50}} \put(-100,0){\vector(1,0){50}}
\put(-117,-7){{\small $-\varrho$}} \put(102,-7){{\small $\varrho$}}
\put(2,55){{\small $\pi/h$ }} \put(2,-59){{\small $-\pi/h$ }} \put(130,-10){
{\small Re($\theta$)}} \put(2,90){{\small Im($\theta$)} } \put(-3.3,48){$%
\bullet$ } \put(-3.3,-53){$\bullet$ } \put(-3.3,20){$\circ$ } \put(-3.3,-20){%
$\circ$ } \put(2,20){ {\small $ \frac{\pi B}{h}$ }} \put(2,-20){ {\small $-%
\frac{\pi B}{h}$ } } \put(80,35){ ${\cal C}_{\varrho }^{+}$ } \put(80,-43){ $%
{\cal C}_{\varrho }^{-}$ } 
\]
\caption{\small The contours 
${\cal C}_{\varrho }^{\pm }$ in the complex $\theta$%
-plane. The bullets $\bullet$ belong to poles resulting from $%
\omega_2(\theta)$ and the open circles $\circ$ to poles of 
$\omega_1(\theta)$.}
\label{Fig1}
\end{figure}

Since  the identity (\ref{funi}) is by no means 
obvious, in particular with regard to (\ref{FTS}), we will now 
provide a rigourous proof of it. Essentially we have to use the 
Fourier transformed version of the relation 
(\ref{rel}) for this purpose. Shifting the Fourier integral into the complex
plane yields 
\begin{eqnarray}
\lim\limits_{\varrho \rightarrow \infty }\oint\limits_{{\cal C}_{\varrho
}^{\pm }}d\theta \varphi _{ij}(\theta )e^{ik\theta } &=&\widetilde{\varphi }%
_{ij}(k)-{\cal P}\int d\theta \varphi _{ij}\left( \theta \pm i\frac{\pi }{h}%
\right) e^{ik(\theta \pm i\pi /h)}  \label{h2} \\
&=&2\pi \delta _{ij}e^{\mp \pi Bk/h}\,-\pi I_{ij}e^{\mp \pi k/h}\,.
\label{h3}
\end{eqnarray}
Here the contours ${\cal C}_{\varrho }^{\pm }$ are depicted in figure \ref
{Fig1}. ${\cal P}$ denotes the Cauchy principal value. The poles inside the
contours are collected by considering (\ref{kernel}) and (\ref{hyprep}).
Obviously coupling constant dependent poles may only result from $\omega
_{1}(\theta )$ and $\omega _{2h-1}(\theta )$, whilst poles directly on the
contours at $\theta =\pm i\pi /h$ may only originate from $\omega
_{2}(\theta )$ and $\omega _{2h-2}(\theta )$. From the statements made in
subsection 4.1.1., it follows that the former blocks may only occur when $%
i=j $ and from equation (\ref{2block}) we infer that the pre-factor of $%
\omega _{2}(\theta )$ is $I_{ij}$. The relevant residues are computed easily
and by noting further that the singularities directly on the contour count
half, we have established (\ref{h3}).

Acting now with $-i$ times the logarithmic derivative on the identity (\ref
{rel}), multiplying with $\exp (ik\theta )$ and integrating thereafter with
respect to $\theta $ we obtain 
\begin{equation}
{\cal P}\int d\theta \left( \varphi _{ij}\left( \theta +i\pi /h\right)
+\varphi _{ij}\left( \theta -i\pi /h\right) \right) e^{ik \theta
}=\sum\limits_{l=1}^{r}I_{il}\widetilde{\varphi }_{lj}(k)\,\,.  \label{h22}
\end{equation}
On the other hand the l.h.s. of (\ref{h22}) may be computed alternatively
from the right hand sides of (\ref{h2}) and (\ref{h3}), such that we obtain 
\begin{equation}
\sum\limits_{l=1}^{r}I_{il}\widetilde{\varphi }_{lj}(k)=2\widetilde{\varphi }%
_{ij}(k)\cosh \frac{\pi }{h}k+2\pi I_{ij}-4\pi \delta _{ij}\cosh \frac{\pi }{%
h}k(1-B)\,\,,  \label{h33}
\end{equation}
and therefore (\ref{funi}).

When we take the limit $B\rightarrow i\infty $, the coupling constant
dependent poles move outside the contours ${\cal C}_{\varrho }^{\pm }$, such
that the term involving $B$ will be absent in (\ref{h33}) and we recover the
result of \cite{TBAZamun}. The weak limit $B\rightarrow 0$ turns equ. (\ref
{h33}) into a trivial identity by noting that with (\ref{funi}) $%
\lim_{B\rightarrow 0}\widetilde{\varphi }_{ij}(k)=2\pi \delta _{ij}$.

As the last quantity which will be important for our considerations we have
to compute the power series expansion of $\widetilde{\varphi }_{ij}(k)$. For
this purpose we make use of the identity 
\begin{equation}
\frac{\sinh (xt)}{\sinh t}=\sum\limits_{n=0}^{\infty }\frac{2^{2n+1}}{(2n+1)!%
}B_{2n+1}\left( \frac{1+x}{2}\right) t^{2n}\,,
\end{equation}
which follows directly from the generating function for the Bernoulli
polynomials $B_{n}(x)$ of degree $n$ (e.g. \cite{Grad}). We may then expand
each building block of the Fourier transformed TBA-kernel (\ref{genexp}) as 
\begin{eqnarray}
\widetilde{\omega }_{x}(k) &=&2\pi \sum\limits_{n=0}^{\infty }\mu
_{x}^{(2n)}k^{2n}\,\,  \label{Kexp} \\
&=&2\pi \sum\limits_{n=0}^{\infty }\left( \nu _{x+1}^{(2n)}+\nu
_{x-1}^{(2n)}-\nu _{x+1-B}^{(2n)}-\nu _{x-1+B}^{(2n)}\right) k^{2n}
\label{mu}
\end{eqnarray}
where the coefficients are $\nu _{x}^{(2n)}=2(2\pi
)^{2n}/(2n+1)!B_{2n+1}(x/(2h))$. Therefore we obtain for the Fourier
coefficients in equation (\ref{Fexp}) 
\begin{equation}
\eta _{ij}^{(0)}=\delta _{ij}-g_{ij}\quad \text{and}\quad \eta _{ij}^{(2n)}=%
\frac{(-1)^{n+1}}{2}\sum\limits_{q=1}^{h}(\lambda _{i}\cdot \sigma
^{q+(c(j)-1)/2}\gamma _{j})\mu _{2q+(c(j)-c(i))/2-1}^{(2n)}\,\;
\end{equation}
for $n=1,2,\ldots $ The second order coefficient is particularly important
with regard to our approximated analytical solution presented in section 3.2
and we will therefore analyze it a bit further. From (\ref{mu}) we obtain $%
\mu _{x}^{(2)}=\pi ^{2}B(B-2)(h-x)/h^{3}$, such that 
\begin{equation}
\eta _{ij}^{(2)}=\frac{1}{2h^{3}}\sum\limits_{q=1}^{h}(\lambda _{i}\cdot
\sigma ^{q+(c(j)-1)/2}\gamma _{j})\pi ^{2}B(B-2)(h-2q-(c(j)-c(i))/2+1)\,\,.
\label{eta}
\end{equation}
We may evaluate the sum over $q$ by noting further that 
\begin{equation}
\qquad \lambda _{i}=\frac{1}{h}\sum\limits_{q=1}^{h}q\sigma
^{q+(c(i)-1)/2}\gamma _{i}\quad \quad \text{and\quad \quad }%
\sum\limits_{q=1}^{h}\lambda _{i}\cdot \sigma ^{q}\gamma _{j}=0\,\,.\text{ }
\end{equation}
The first identity follows by inverting the relation $\gamma _{i}=(1-\sigma
^{-1})\sigma ^{(1-c(i))/2}\lambda _{i}$ \cite{FO} and the second by
computing the geometric series. Therefore (\ref{eta}) becomes 
\begin{equation}
\eta _{ij}^{(2)}=\frac{\pi ^{2}B(2-B)}{h^{2}}\lambda _{i}\cdot \lambda _{j}=%
\frac{\pi ^{2}B(2-B)}{h^{2}}K_{ij}^{-1}\quad .  \label{eta2ij}
\end{equation}
The latter formula follows immediately by recalling that $\lambda
_{i}=\sum_{j=1}^{r}K_{ij}^{-1}\alpha _{j}$. It has the virtue that it is
directly applicable since it involves only quantities which may be
effortlessly extracted from Lie algebraic tables. At last we may compute the
quantity $\eta _{i}^{(2)}=\sum_{j=1}^{r}\eta _{ij}^{(2)}$ which occurs in
our approximated analytical approach (\ref{Soll}). Obviously we obtain from (%
\ref{eta2ij}) 
\begin{equation}
\eta _{i}^{(2)}=\frac{\pi ^{2}B(2-B)}{h^{2}}\lambda _{i}\cdot \rho \,\,,
\end{equation}
where $\rho =\sum_{i=1}^{r}\lambda _{i}=1/2\sum_{\alpha >0}\alpha $ is the
Weyl vector. The inner product of $\lambda _{i}\cdot \rho $ may be related
to a universal quantity, namely the index of the fundamental representation $%
\lambda _{i}$%
\begin{equation}
x_{\lambda _{i}}=\frac{\text{dim}\lambda _{i}}{2\text{dim{\bf g}}}(2\rho
+\lambda _{i})\cdot \lambda _{i}\,\,,
\end{equation}
such that we finally obtain 
\begin{equation}
\eta _{i}^{(2)}=\frac{\pi ^{2}B(2-B)}{2h^{2}}\left( \frac{2x_{\lambda _{i}}%
\text{dim{\bf g}}}{\text{dim}\lambda _{i}}-K_{ii}^{-1}\right) \,\,.
\end{equation}
Hence we are also able to evaluate $\eta _{i}^{(2)}$ in a universal manner.
We may proceed further and also compute 
\begin{equation}
\sum_{i=1}^{r}\eta _{i}^{(2)}=\frac{\pi ^{2}B(2-B)}{h^{2}}\rho ^{2}=\frac{%
\pi ^{2}B(2-B)Q_{\psi }\text{dim{\bf g}}}{24h^{2}}=\frac{\pi ^{2}B(2-B)r(h+1)%
}{12h}.  \label{sumeta}
\end{equation}
Here we used the Freudenthal-deVries strange formula $\rho ^{2}=Q_{\psi }$dim%
{\bf g}$/24$ (see e.g. \cite{GO}), the fact that for simply laced algebras
the eigenvalue of the quadratic Casimir operator is $Q_{\psi }=2h$ and that
we have dim{\bf g}$=r(h+1)$ for the dimension of the Lie algebra {\bf g}.

\subsubsection{Non-simply laced Lie algebras}

For the non-simply laced cases we may proceed similarly, just now we are
lacking the universality of the previous subsection. Using the generalized
blocks (\ref{nonblocks}) instead of (\ref{block}) we derive 
\begin{eqnarray}
\omega _{x,y}(\theta ) &=&-i\frac{d}{d\theta }\ln \left[ x,y\right] _{\theta
}=-\frac{1}{\pi }\int\limits_{0}^{\infty }\frac{dt}{\sinh t}%
\,\,g_{x,y,B}(t)\cos \left( \frac{\theta t}{\pi }\right) , \\
&=&\gamma _{x-yB-1}(\theta )+\gamma _{x+yB+1}(\theta )-\gamma
_{x+yB+B-1}(\theta )-\gamma _{x-yB-B+1}(\theta )\,.
\end{eqnarray}
The Fourier transformed of $\omega _{x,y}(\theta )$ reads 
\begin{eqnarray}
\widetilde{\omega }_{x,y}(k) &=&-\frac{\pi \,g_{x,y,B}(\pi k)}{\sinh (\pi k)}%
\, \\
&=&2\pi \sum\limits_{n=0}^{\infty }\mu _{x,y}^{(2n)}k^{2n}\,\, \\
&=&2\pi \sum\limits_{n=0}^{\infty }\left( \nu _{x+1+yB}^{(2n)}+\nu
_{x-1-yB}^{(2n)}-\nu _{x+1-B-yB}^{(2n)}-\nu _{x-1+B+yB}^{(2n)}\right)
k^{2n}\,.
\end{eqnarray}
In particular 
\begin{equation}
\mu _{x,y}^{(2)}=\frac{B\pi ^{2}(B-2)(H-x)(1-2y)}{H^{3}}\,\quad
\end{equation}
is important for our purposes.

\subsection{The TBA-equations}

Having computed the TBA-kernels, it appears at first sight that, apart from
a simple substitution, there is not much more to be said about the form of
the TBA-equations. However, Zamolodchikov observed in \cite{TBAZamun} that
once the Fourier transformed  TBA-kernels admit a certain
representation (in our situation this is (\ref{funi})), the TBA-equations
may be cast into a very universal form.

We will exploit now the universal features derived in the preceding section.
In order to keep the notation simple, we commence by choosing the
statistical interaction to be fermionic at first. The generalization to
generic Haldane statistics is straightforward thereafter. First of all, we
Fourier transform the TBA-equations in the variant (\ref{eps}) 
\begin{equation}
2\pi \widetilde{\xi }_{i}(k,r)=\sum\limits_{j=1}^{r}\widetilde{\varphi }%
_{ij}(k)\widetilde{L}_{j}(k,r)\,.  \label{FTTBAF}
\end{equation}
For the reason that the Fourier transforms of $rm_{i}\cosh \theta $ and $%
\varepsilon _{i}(\theta ,r)$ do not exist separately, we introduced here the
quantity $\xi _{i}(\theta ,r)=rm_{i}\cosh \theta $ $-\varepsilon _{i}(\theta
,r)$. This should be kept in mind before transforming back to the original
variables. Substituting now the general form of $\widetilde{\varphi }%
_{ij}(k) $ from (\ref{funi}) into equation (\ref{FTTBAF}) and taking the
inverse Fourier transformation thereafter yields 
\begin{equation}
\xi _{i}(\theta ,r)=\sum\limits_{j=1}^{r}\left( I_{ij}(\xi
_{j}-L_{j})*\Omega _{h}\right) (\theta ,r)+(L_{i}*\Omega _{h,B})(\theta
,r)\,\,.  \label{uniTBA}
\end{equation}
Here we introduced the universal kernels 
\begin{equation}
\Omega _{h}(\theta )=\frac{h}{2\cosh \frac{h}{2}\theta }\quad \text{and
\quad }\Omega _{h,B}(\theta )=\frac{2h\sin \frac{\pi }{2}B\cosh \frac{h}{2}%
\theta }{\cosh h\theta -\cos \pi B}\,\,.
\end{equation}
Neglecting the coupling constant dependent term involving $\Omega
_{h,B}(\theta )$, i.e. taking the limit $B\rightarrow i\infty $, the
identities (\ref{uniTBA}) coincide precisely with the equations (7) in \cite
{TBAZamun}. This is to be expected since the latter equations describes the
system which dynamically interacts via the minimal part of (\ref{S}). Note
further that the strong-weak duality is still preserved, i.e. $\Omega
_{h,B}(\theta )=\Omega _{h,2-B}(\theta )$.

We may now easily derive the generalized form of (\ref{uniTBA}) for generic
types of Haldane statistics. Recalling that for the derivation of (\ref{eps}%
) all the $L-$functions were parameterized by ${\cal L}_{i}(\theta ,r)=\ln
\left( 1+e^{-\varepsilon _{i}(\theta ,r)}\right) $ and the statistical
interaction by $g_{ij}=\delta _{ij}-g_{ij}^{\prime }$, the Fourier
transformation reads 
\begin{equation}
\widetilde{\xi }_{i}(k,r)=\sum\limits_{j=1}^{r}\left[ g_{ij}^{\prime }\,%
\widetilde{{\cal L}}_{j}(k,r)+\widetilde{\varphi }_{ij}(k)\widetilde{{\cal L}%
}_{j}(k,r)/2\pi \right] \,.  \label{FTTBA}
\end{equation}
In the same manner as for the fermionic case we derive 
\begin{equation}
\xi _{i}(\theta ,r)=\sum\limits_{j=1}^{r}\left[ I_{ij}(\xi
_{j}-\sum\limits_{l=1}^{r}(\delta _{jl}+g_{jl}^{\prime }){\cal L}%
_{l})*\Omega _{h}+g_{ij}^{\prime }{\cal L}_{j}\right] (\theta ,r)+({\cal L}%
_{i}*\Omega _{h,B})(\theta ,r)\,\,.  \label{uniTBAH}
\end{equation}
Of course we recover (\ref{uniTBA}) from (\ref{uniTBAH}) for $g_{ij}^{\prime
}=0$.

As a final consistency check we carry out the limit to the free theory, that
is $B\rightarrow 0$. We have 
\begin{eqnarray}
\lim\limits_{B\rightarrow 0}({\cal L}_{i}*\Omega _{h,B})(\theta ,r)\,
&=&\lim\limits_{B\rightarrow 0}\frac{1}{2}\int\limits_{-\infty }^{\infty
}d\theta ^{\prime }\frac{hB\cosh \frac{h}{2}\theta ^{\prime }}{\cosh h\theta
^{\prime }-1+\pi ^{2}B^{2}/2}{\cal L}_{i}(\theta -\theta ^{\prime },r) \\
&=&\int\limits_{-\infty }^{\infty }dt\lim\limits_{B\rightarrow 0}\frac{1}{%
\pi }\frac{B}{t^{2}+B^{2}}{\cal L}_{i}\left( \theta -\frac{2}{h}\func{arsinh}%
\frac{t\pi }{2},r\right) \\
&=&{\cal L}_{i}(\theta ,r)\,\,.
\end{eqnarray}
In the last equality we employed the well known representation for the delta
function $\pi \delta (x)=\lim_{\varepsilon \rightarrow 0}\varepsilon
/(\varepsilon ^{2}+x^{2})$. This means that in the extreme weak limit the
TBA-equations (\ref{uniTBAH}) is solved by 
\begin{equation}
\varepsilon _{i}(\theta ,r)=rm_{i}\cosh \theta -\sum\limits_{j=1}^{r}(\delta
_{ij}+g_{ij}^{\prime })L_{j}(\theta ,r)\,\,.  \label{B0}
\end{equation}
That (\ref{B0}) are indeed {\it the} TBA-equations may also be seen by
taking this limes directly in (\ref{eps}) and noting that $%
\lim_{B\rightarrow 0}\varphi _{ij}(\theta )=\delta _{ij}(\theta )$. In this
way we have ``switched off'' the dynamical interaction and obtained a system
which interacts purely by statistics. In particular for fermionic type of
statistics ($g_{ij}^{\prime }=0$) or for $g_{ij}^{\prime }=-\delta _{ij}$
these equations describe $r$ free bosons or $r$ free fermions, respectively.
Taking $g_{ij}^{\prime }$ to be diagonal, the equations (\ref{B0}) are
equivalent to a system whose S-matrix corresponds to the one of the
Calogero-Sutherland model and whose statistical interaction is taken to be
of fermionic type \cite{BF}.

We finish this subsection by noting that $\sum_{j}I_{ij}m_{j}=2m_{i}\cos
(\pi /h)$ \cite{FLO}, from which follows 
\begin{equation}
rm_{i}\cosh \theta =r\sum\limits_{j=1}^{r}I_{ij}m_{j}\left( \cosh *\Omega
_{h}\right) (\theta ,r)\,\,,
\end{equation}
such that the entire $r$-dependence and the mass spectrum in (\ref{uniTBA})
and (\ref{uniTBAH}) may be eliminated. We obtain 
\begin{equation}
\varepsilon _{i}(\theta ,r)=\sum\limits_{j=1}^{r}\left[ I_{ij}(\varepsilon
_{j}+\sum\limits_{l=1}^{r}(\delta _{jl}+g_{jl}^{\prime }){\cal L}%
_{l})*\Omega _{h}-g_{ij}^{\prime }{\cal L}_{j}\right] (\theta ,r)-({\cal L}%
_{i}*\Omega _{h,B})(\theta ,r)\,\,
\end{equation}
instead. One may solve these equations and re-introduce the $r$-dependence
thereafter by means of the asymptotic behaviour. This is similar to the
situation in the next subsection, in which the equations are even further
simplified to a set of functional equations.

\subsection{Y-systems}

An alternative analytical approach is provided by exploiting properties of
what is often referred to as Y-systems \cite{TBAZamun}. In order to derive
these equations, we invoke once more the convolution theorem on the
convolution in equation (\ref{eps}) and compute thereafter the sum of (\ref
{eps}) at $\theta +i\pi /h$ and $\theta -i\pi /h$ minus $\sum_{j}I_{ij}$
times (\ref{eps}).Using once more the fact that $\sum_{j}I_{ij}m_{j}=2m_{i}%
\cos (\pi /h)$, the terms involving $m_{i}$ cancel. Employing thereafter the
identity (\ref{h33}) and transforming back into the original $\theta -$space
we obtain, upon the introduction of the new variable $Y_{i}\left( \theta
\right) =\exp (-\varepsilon _{i}(\theta ))$,

\begin{eqnarray}
\begin{array}{c}
Y_{i}\left( {\theta +}\frac{i\pi }{h}\right) Y_{i}\left( {\theta -}\frac{%
i\pi }{h}\right) =\left[ 1+Y_{i}\left( \theta +\frac{i\pi }{h}(1-B)\right)
\right] \left[ 1+Y_{i}\left( \theta -\frac{i\pi }{h}(1-B)\right) \right]
\qquad
\end{array}
\nonumber \\ 
\begin{array}{c}
\times \prod\limits_{j=1}^{r}\left( 1+Y_{j}^{-1}\left( \theta \right)
\right) ^{-I_{ji}}\left( \left[ 1+Y_{j}\left( {\theta +}\frac{i\pi }{h}%
\right) \right] \left[ 1+Y_{j}\left( {\theta -}\frac{i\pi }{h}\right)
\right] \prod\limits_{l=1}^{r}\left[ 1+Y_{l}\left( \theta \right) \right]
^{-I_{jl}}\right) ^{g_{ij}^{\prime }}.
\end{array}
\nonumber
\end{eqnarray}
For fermionic statistics, i.e. $g_{ij}^{\prime }=0$  we recover the
systems stated in \cite{ZamoR} ($h=2$) and \cite{MM}, 
when we chose the ``roaming parameter''
therein such that $S(\theta )$ equals (\ref{S}). The
strong-weak duality is evidently still preserved. The Y-systems
corresponding to the ``minimal theories'' are obtained by taking the limit $%
B\rightarrow i\infty $, by noting that $\lim_{\theta \rightarrow \infty
}Y_{i}\left( {\theta }\right) =0$. The latter limes follows from the
TBA-equations together with the explicit form of the TBA-kernel, such that
for large $\theta $ we have $\varepsilon _{i}(\theta )\sim rm_{i}\cosh
\theta $. For fermionic type of statistics the functional equations obtained
in this limit coincide with those found in \cite{TBAZamun}. As is to be
expected from (\ref{B0}), we obtain for $g_{ij}^{\prime }=0$ (for $h=0$ this
holds in general) in the extremely weak coupling limit 
\begin{equation}
Y_{i}\left( {\theta +}2\pi i\right) =Y_{i}\left( {\theta }\right) ,\qquad 
\text{ }B\rightarrow 0,2\,\,.  \label{YB0}
\end{equation}
To derive these periodicities directly from the functional equations is
rather cumbersome once the algebras are more complicated and we do not have
a general case independent proof. This is similar to the situation in which
the system involved is the minimal part of (\ref{S}) \cite{TBAZamun}. We
leave it for future investigations to exploit the functional equations
further.

\subsection{Scaling functions}

As already stated, solving the TBA-equations and the subsequent evaluation
of (\ref{c(r)}) leads to expressions of the scaling function $c(r)$ with the
inverse temperature $r$ times a mass scale as the scaling parameter. We may
carry out this task numerically. Alternatively, assembling our results from
section 3 and the analysis of the TBA-kernel, we can write down a universal
formula for the scaling function of the system in which the statistical
interaction is of fermionic type and in which the dynamical interaction is
governed by the scattering matrix of affine Toda field theories related to
simply laced Lie algebras, up to the first leading order in the ultraviolet
regime. From (\ref{cappp}) and (\ref{sumeta}) it follows under the
assumption that the constants $\beta _{i}$ are the same for all particle
types 
\begin{equation}
c^{{\bf g}}(r)=\text{rank{\bf g}}\left( 1-\frac{\pi ^{2}B(2-B)(h+1)}{%
4h(\beta -\ln (r/2))^{2}}\right) \quad .  \label{universal}
\end{equation}
We will now compute step by step several scaling functions for affine Toda
field theories in which we specify some concrete algebras. We gradually
increase the complexity of the models and focus on different aspects.

\subsubsection{$A_{1}^{(1)}$-ATFT (Sinh-Gordon)}

\noindent The Sinh-Gordon theory is one of the simplest quantum field
theories in 1+1 dimensions and therefore the ideal starting point to
concretely test general statements made in this context. Semi-classical
investigations of a TBA-type nature concerning this model may already be
found in \cite{BPT}. The model contains only one stable particle which does
not fuse and its scattering matrix simply reads 
\begin{equation}
S^{SG}\left( \theta \right) =\left\{ 1\right\} _{\theta }=\frac{\tanh \frac{1%
}{2}\left( \theta -\frac{i\pi }{2}B\right) }{\tanh \frac{1}{2}\left( \theta +%
\frac{i\pi }{2}B\right) }\quad .
\end{equation}
For a system whose dynamics is described by this S-matrix and whose
statistical interaction is of general Haldane type, the TBA-equation (\ref
{HTBA}) reads 
\begin{equation}
rm\cosh \theta +\ln (1-e^{-L(\theta ,r)})=\frac{1}{2\pi }\int\limits_{-%
\infty }^{\infty }d\theta ^{\prime }\varphi ^{SG}\left( \theta ^{\prime
}\right) L(\theta -\theta ^{\prime },r)-gL(\theta ,r)  \label{TBASG}
\end{equation}
with 
\begin{equation}
\varphi ^{SG}\left( \theta \right) =\Omega _{2,B}(\theta )=\frac{4\sin (\pi
B/2)\cosh \theta }{\cosh 2\theta -\cos \pi B}\,.
\end{equation}
We may convince ourselves that (\ref{TBASG}) can be obtained from (\ref{HTBA}%
) and the direct computation of the kernel as well as from (\ref{uniTBAH})
by noting that the incidence matrix of course vanishes for the $A_{1}^{(1)}$
related Dynkin diagram.

\vspace*{-1.7cm}
\begin{center}
\includegraphics[width=11cm,height=16cm,angle=-90]{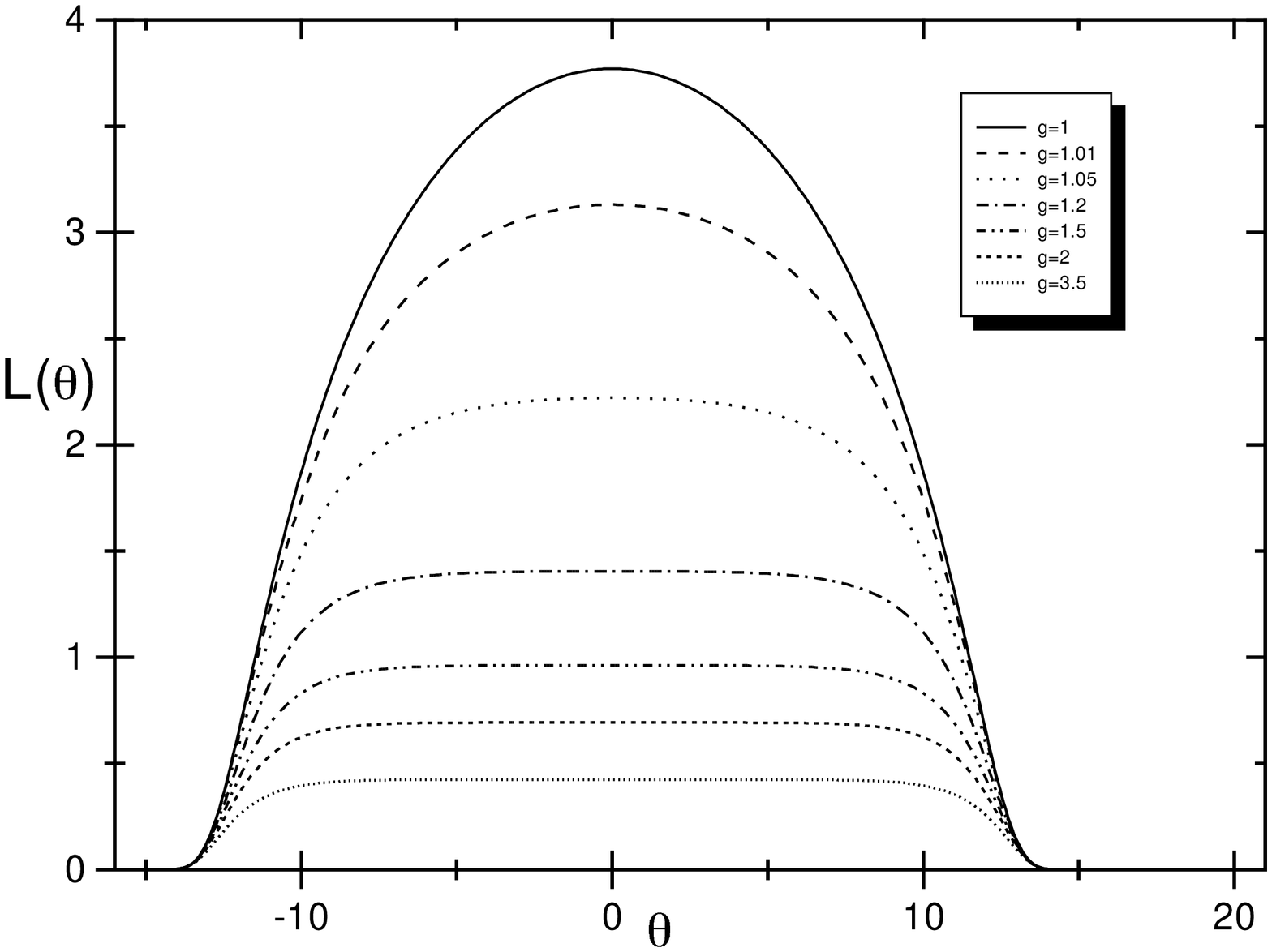}
\end{center}
\vspace*{-1cm} 
{\small Figure 2: Numerical solution for  the Sinh-Gordon related TBA-equations
for different statistical  interactions $g$ and fixed values of the effective 
coupling $B=0.4$ and $r = 10^{-5}$.}
\vspace*{1.2mm}

The weak coupling limit $B\rightarrow 0$ is read off from (\ref{B0}) 
\begin{equation}
rm\cosh \theta +\ln (1-e^{-L(\theta ,r)})+gL(\theta ,r)=L(\theta ,r)\quad .
\label{40}
\end{equation}

The other limes which may be computed in the standard way is the limes $%
r\rightarrow 0,$ leading to the constant TBA-equation 
\begin{equation}
\ln \left( 1-e^{-L(0,0)}\right) =(1-g)L(0,0)\quad .  \label{SGex}
\end{equation}
This is obviously compatible with equation (\ref{40}) and indicates that the
extreme ultraviolet behaviour is independent of the dynamics and is purely
governed by the statistics. In particular we obtain for fermionic type of
statistics $L(0,0)\rightarrow \infty $ as a solution of (\ref{SGex}), such
that with (\ref{cex}) the effective central charge turns out to be one. In
general (\ref{SGex}) yields $0\leq c_{\text{eff}}\leq 1$ for $g\geq 1$. We
also observe that due to the fact that $L(0,0)\geq 0$, equation (\ref{SGex})
does not possess any physical solutions at all for $g<1$, which means in
particular that there are no solutions for the statistical interaction of
bosonic type.

We shall now turn to the full solution of the TBA-equation (\ref{TBASG}).
First of all we solve this equation numerically. For $g\geq 1$ we observe in
figure 2 the typical behaviour, known from the analysis of the minimal part
of the S-matrices \cite{TBAZam}, that the $L$-function is essentially
constant between $\pm 2\ln (r/2)$. As expected from the previous discussion,
also the numerical analysis does not yield any solution for $g<1$. For a
more detailed discussion concerning the nature of the numerical procedure
and in particular the question of convergence and uniqueness of the
solutions we refer to section 5.

We shall now compare our numerical results with our analytic approximations
of section 3. The general formulae (\ref{genexp}) or (\ref{funi}) yields for
the Fourier transform of the TBA-kernel 
\begin{equation}
\widetilde{\Phi }^{SG}(k)=
\frac{2\pi \cosh \left( \frac{k\pi }{2}-\frac{Bk\pi }{2}\right)}
{\cosh \left( \frac{k\pi }{2}\right) }-2\pi g\quad .
\label{SGtrans}
\end{equation}

For the reasons mentioned in section 3.2, we will now restrict ourselves to
fermionic type of statistics, such that 
\begin{equation}
\eta ^{(0)}=0\qquad \text{and}\qquad \eta ^{(2)}=\frac{\pi ^{2}B(2-B)}{8}\,.
\end{equation}
We obtain a fairly good agreement between the numerical solution of (\ref
{HTBA}) for the function $L(\theta ,r)$ and the approximated analytical
solution (\ref{Lappp}), if we choose the constants in (\ref{Soll}) to be 
\begin{equation}
\beta =\ln (B(2-B)2^{1+B(2-B)})\qquad \text{and\qquad }\alpha =\frac{\pi }{%
2(\beta -\ln (r/2))}\,\,,  \label{Konstanten}
\end{equation}
such that 
\begin{equation}
L(\theta ,r)=\ln \left( \cos ^{2}\left( \alpha \theta \right) \right) -\ln
[B(2-B)(\pi \alpha /2)^{2}]\,\,.  \label{appSG}
\end{equation}

\vspace{-1.9cm}
\begin{center}
\includegraphics[width=11cm,height=16cm]{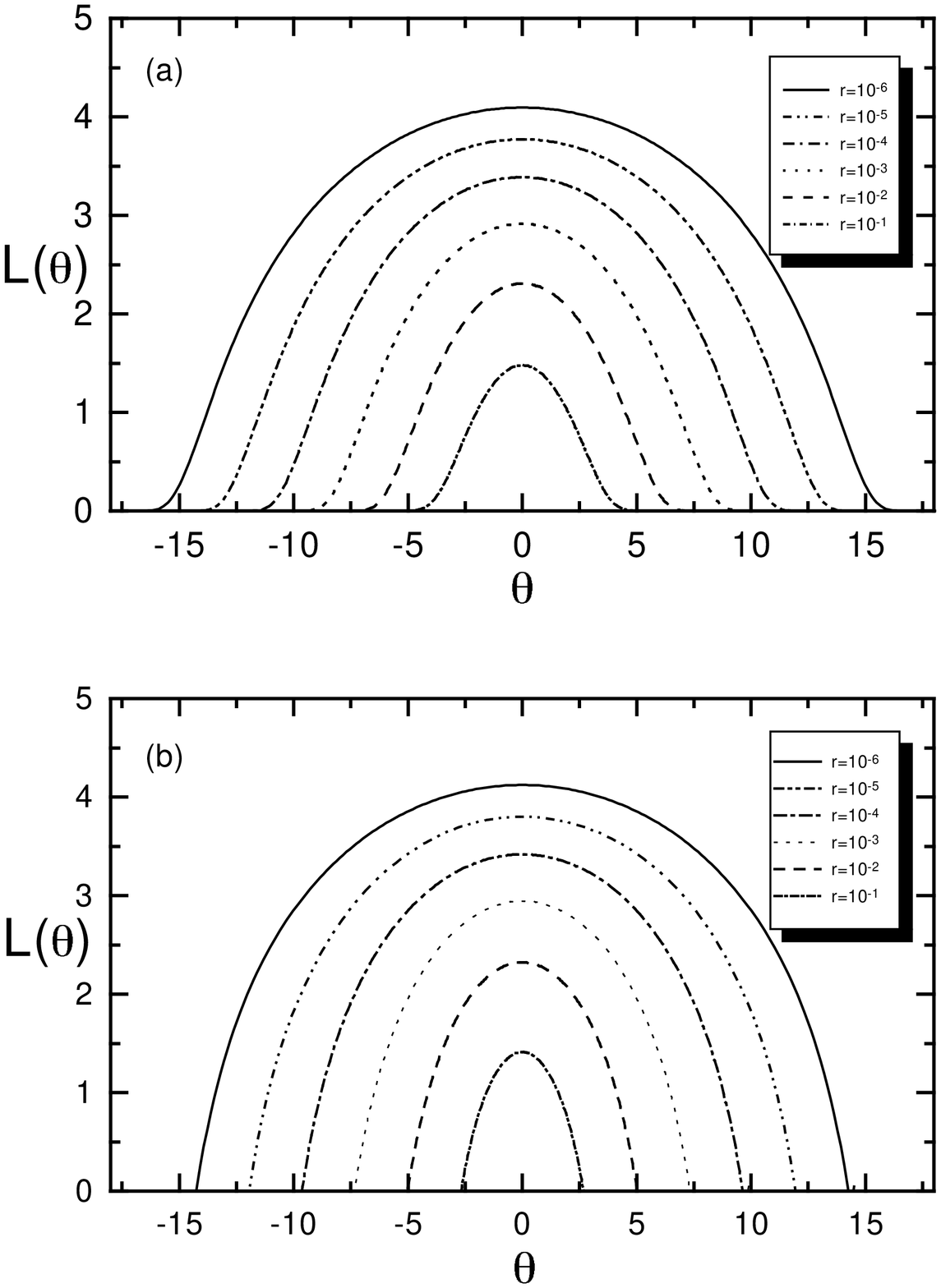}
\end{center}
\vspace{-1.5cm}
{\small Figure 3: Numerical solution (a) 
versus approximated analytical solution (b) for the
Sinh-Gordon related TBA-equation for various values of $r$ 
and fixed effective coupling $B=0.4$.}
\vspace*{1.2mm}

Figure 3 shows that the $r$ dependence of $L(\theta ,r)$ is captured very
well by the approximated analytical solution (\ref{appSG}). As expected from
the nature of the assumptions made in the derivation, (\ref{Lappp}) becomes
relatively poor when $L(\theta ,r)$ is small. With regard to the aim of
these computations, that is the evaluation of the scaling function $c(r)$,
precisely in this region the error is negligible as is seen from 
(\ref{c(r)}). Notice also 
that the assumptions (\ref{slowvary}) are justified in
retrospective by the numerical solutions. The quality of the approximation
may also be seen by comparing tables 1-3 for large values of $L(\theta ,r).$
\vspace{-1.5cm}
\begin{center}
\includegraphics[width=11cm,height=16cm,angle=-90]{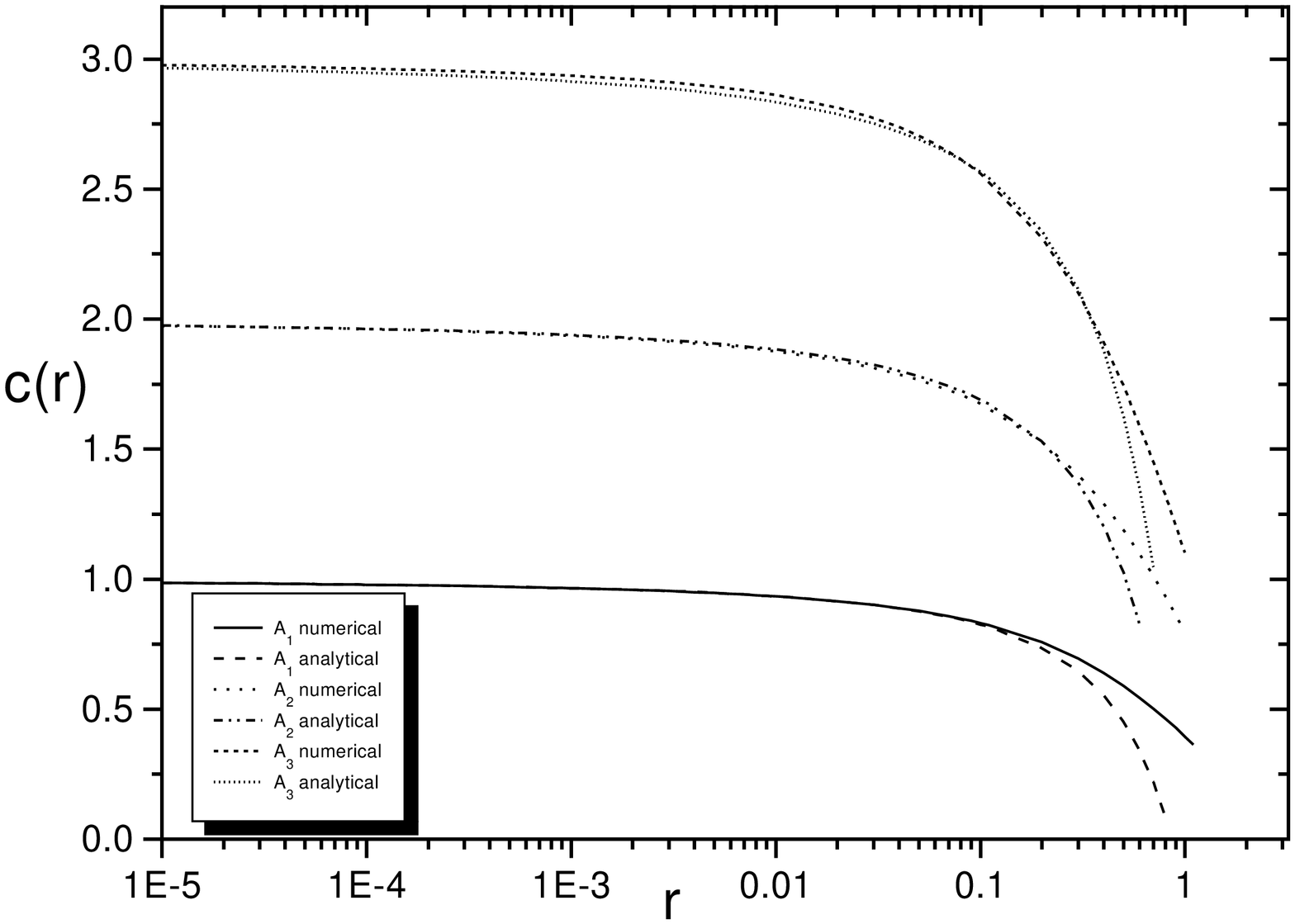}
\end{center}
\vspace{-0.95cm}
\vspace*{0.5mm}
{\small Figure 4: Numerical solution  
versus approximated analytical solution for the
scaling function
of the $A_1^{(1)},  A_2^{(1)} $ and $A_3^{(1)}$-affine Toda field theory
 related TBA-systems with fixed effective coupling $B=0.4$.}

Finally we compute the scaling function, which acquires in this
approximation the form 
\begin{equation}
c^{SG}(r)=1-\frac{3\,\,\pi ^{2}B(2-B)}{8(\beta -\ln (r/2))^{2}}\quad .
\end{equation}
Also this function is well approximated in the ultraviolet regime ($r<1$) as
is seen from figure 4 and the tables 1-3.

\begin{center}
\begin{tabular}{|l|c|c|c|c|}
\hline\hline
$r$ & $L(0,r)$/n. & $L(0,r)$/a. & $c(r)$/ n. & $c(r)$/a. \\ \hline\hline
$10^{-1}$ & 1.845 & 2.135 & 0.8767 & 0.9281 \\ \hline
$10^{-2}$ & 2.957 & 3.238 & 0.9648 & 0.9762 \\ \hline
$10^{-3}$ & 3.722 & 3.945 & 0.9845 & 0.9882 \\ \hline
$10^{-4}$ & 4.286 & 4.467 & 0.9914 & 0.9930 \\ \hline
$10^{-5}$ & 4.740 & 4.880 & 0.9949 & 0.9954 \\ \hline
$10^{-6}$ & 5.153 & 5.222 & 0.9975 & 0.9967 \\ \hline\hline
\end{tabular}
\end{center}
\vspace*{1mm} 
{\small Table 1: Numerical solution (n.) 
versus approximated analytical solution
(a.) for the Sinh-Gordon related TBA-system with fixed effective coupling $%
B=0.1.$ }

\begin{center}
\begin{tabular}{|l|c|c|c|c|}
\hline\hline
$r$ & $L(0)$/n. & $L(0)$/a. & $c(r)$/ n. & $c(r)$/a. \\ \hline\hline
$10^{-1}$ & 1.478 & 1.249 & 0.8311 & 0.8257 \\ \hline
$10^{-2}$ & 2.307 & 2.220 & 0.9339 & 0.9340 \\ \hline
$10^{-3}$ & 2.915 & 2.870 & 0.9654 & 0.9655 \\ \hline
$10^{-4}$ & 3.387 & 3.361 & 0.9789 & 0.9789 \\ \hline
$10^{-5}$ & 3.771 & 3.754 & 0.9857 & 0.9858 \\ \hline
$10^{-6}$ & 4.094 & 4.082 & 0.9897 & 0.9897 \\ \hline\hline
\end{tabular}
\end{center}
\vspace*{1mm} 
{\small Table 2: Numerical solution (n.) 
versus approximated analytical solution
(a.) for the Sinh-Gordon related TBA-system with fixed effective coupling $%
B=0.4.$}

\begin{center}
\begin{tabular}{|l|c|c|c|c|}
\hline\hline
$r$ & $L(0,r)$/n. & $L(0,r)$/a. & $c(r)$/ n. & $c(r)$/a. \\ \hline\hline
$10^{-1}$ & 1.335 & 0.811 & 0.8070 & 0.7297 \\ \hline
$10^{-2}$ & 2.074 & 1.782 & 0.9165 & 0.8977 \\ \hline
$10^{-3}$ & 2.631 & 2.433 & 0.9540 & 0.9466 \\ \hline
$10^{-4}$ & 3.072 & 2.923 & 0.9709 & 0.9673 \\ \hline
$10^{-5}$ & 3.435 & 3.317 & 0.9800 & 0.9780 \\ \hline
$10^{-6}$ & 3.743 & 3.646 & 0.9853 & 0.9841 \\ \hline\hline
\end{tabular}
\end{center}
\vspace*{1mm} 
{\small Table 3: Numerical solution (n.) 
versus approximated analytical solution
(a.) for the Sinh-Gordon related TBA-system with fixed effective coupling $%
B=0.9.$}

\subsubsection{$A_{2}^{(1)}$-ATFT/$A_{2}^{(2)}$-ATFT (Bullough-Dodd)}

In comparison to the Sinh-Gordon model the next complication arises when we
allow the particle in the system to fuse. This case will arise when we
consider the $A_{2}^{(1)}$- and $A_{2}^{(2)}$-ATFT, which are known to be
closely related. The former contains two particles which are conjugate to
each other, whereas the latter contains only one particle which is the bound
state of itself. With regard to the TBA-analysis these models may be treated
on the same footing. The corresponding two particle scattering matrices read

\begin{equation}
S_{11}^{A_{2}^{(1)}}\left( \theta \right) =S_{22}^{A_{2}^{(1)}}\left( \theta
\right) =\left\{ 1\right\} _{\theta }\qquad S_{12}^{A_{2}^{(1)}}\left(
\theta \right) =\left\{ 2\right\} _{\theta }\qquad S^{A_{2}^{(2)}}\left(
\theta \right) =\left\{ 1\right\} _{\theta }\left\{ 2\right\} _{\theta }\,.
\end{equation}
Due to the fact that $S^{A_{2}^{(2)}}\left( \theta \right)
=S_{11}^{A_{2}^{(1)}}\left( \theta \right) S_{12}^{A_{2}^{(1)}}\left( \theta
\right) $, the TBA-equation of the $A_{2}^{(2)}$-theory equals the two
TBA-equations of the $A_{2}^{(1)}$-theory under the natural assumptions that 
$L_{1}(\theta )=L_{2}(\theta )$ and $g_{ij}=g_{ji}$. The common TBA-kernel
without the statistics factor is computed to 
\begin{equation}
\varphi ^{BD}(\theta )=\frac{-4\sqrt{3}\cosh \theta }{2\cosh (2\theta )+1}+%
\frac{4\cosh \theta \sin ((1+B)\pi /3)}{\cosh (2\theta )-\cos ((1+B)2\pi /3)}%
+\frac{4\cosh \theta \sin (B\pi /3)}{\cosh (2\theta )-\cos (2B\pi /3)}\,.
\end{equation}
We may then write the TBA-equation either in its variant (\ref{HTBA}) as 
\begin{equation}
rm\cosh \theta +\ln (1-e^{-L(\theta ,r)})=\frac{1}{2\pi }\int\limits_{-%
\infty }^{\infty }d\theta ^{\prime }\varphi ^{BD}\left( \theta ^{\prime
}\right) L(\theta -\theta ^{\prime },r)-gL(\theta ,r)  \label{85}
\end{equation}
or in its universal form (\ref{uniTBAH}) as 
\begin{equation}
\xi (\theta ,r)=\left[ (\xi -(1+g^{\prime }){\cal L})*\Omega _{3}\right]
(\theta ,r)+g^{\prime }{\cal L}(\theta ,r)\,+({\cal L}_{i}*\Omega
_{h,B})(\theta ,r)\,.
\end{equation}
We used the abbreviations $g=g_{11}+g_{12}$ and $g^{\prime }=g_{11}^{\prime
}+g_{12}^{\prime }$. The Fourier transformed TBA-kernel is computed from (%
\ref{genexp}) or (\ref{funi})

\begin{eqnarray}
\qquad \widetilde{\Phi }^{A_{2}^{(2)}}(k) &=&\widetilde{\Phi }%
_{11}^{A_{2}^{(1)}}(k)+\widetilde{\Phi }_{12}^{A_{2}^{(1)}}(k)=\widetilde{%
\Phi }_{21}^{A_{2}^{(1)}}(k)+\widetilde{\Phi }_{22}^{A_{2}^{(1)}}(k) \\
&=&\frac{4\pi \left( \sinh \left( \frac{\pi k}{3}\right) +\sinh \left( \frac{%
2\pi k}{3}\right) \right) \left( \cosh \left( \frac{k\pi }{3}-\frac{Bk\pi }{3%
}\right) -\frac{1}{2}\right) }{\sinh (\pi k)}\,-2\pi g\,.\quad \quad
\end{eqnarray}
We restrict now to fermionic type of statistics and chose the constants in
our analytic solution (\ref{Soll}) in the same way as for the Sinh-Gordon
model, i.e. 
\begin{equation}
\beta =\beta _{1}=\beta _{2}=\ln (B(2-B)2^{1+B(2-B)})\quad \text{and\quad }%
\alpha =\alpha _{1}=\alpha _{2}=\frac{\pi }{2(\beta -\ln (r/2))}\,\,,
\end{equation}
such that the approximated function for $L(\theta ,r)$ equals (\ref{appSG}).
The scaling function up to the first leading order (\ref{cappp}) becomes now 
\begin{equation}
c^{A_{2}^{(1)}}(r)=2c^{A_{2}^{(2)}}(r)=2-\frac{2\,\,\pi ^{2}B(2-B)}{3(\beta
-\ln (r/2))^{2}}\quad .
\end{equation}
Solving (\ref{85}) numerically we obtain qualitatively the same kind of
agreement with the approximated analytical solution as for the Sinh-Gordon
model, see figure 4.

\subsubsection{$A_{3}^{(1)}$-ATFT}

The simplest model which involves two inequivalent TBA-equations coupled to
each other is the $A_{3}^{(1)}$-affine Toda field theory. In our conventions
the two-particle scattering matrices for this theory read 
\begin{eqnarray}
&&S_{11}^{A_{3}^{(1)}}\left( \theta \right) =S_{33}^{A_{3}^{(1)}}\left(
\theta \right) =\left\{ 1\right\} _{\theta },\qquad
S_{13}^{A_{3}^{(1)}}\left( \theta \right) =\left\{ 3\right\} _{\theta
},\qquad  \label{S1} \\
&&S_{12}^{A_{3}^{(1)}}\left( \theta \right) =S_{23}^{A_{3}^{(1)}}\left(
\theta \right) =\left\{ 2\right\} _{\theta },\qquad
S_{22}^{A_{3}^{(1)}}\left( \theta \right) =\left\{ 1\right\} _{\theta
}\left\{ 3\right\} _{\theta }\,.  \label{S22}
\end{eqnarray}
Particle $1$ is the anti-particle of $3$ and particle $2$ is self-conjugate.
The masses are $m_{1}=m_{3}=m/\sqrt{2}$ and $m_{2}=m$. In the direct channel
the fusings $11\rightarrow 2$ and $33\rightarrow 2$ are possible. Under the
assumption that $\varepsilon _{1}(\theta ,r)=\varepsilon _{3}(\theta ,r)$,
the TBA-equations in the variant (\ref{eps}) now read 
\begin{eqnarray}
\varepsilon _{1}(\theta ,r) &=&rm/\sqrt{2}\cosh \theta -\left( \varphi
_{22}^{A_{3}^{(2)}}*{\cal L}_{1}+\varphi _{12}^{A_{3}^{(2)}}*{\cal L}%
_{2}\right) (\theta ,r)  \nonumber \\
&&-(g_{11}^{\prime }+g_{13}^{\prime }){\cal L}_{1}(\theta ,r)-g_{12}^{\prime
}{\cal L}_{2}(\theta ,r),  \label{TBA12} \\
\varepsilon _{2}(\theta ,r) &=&rm\cosh \theta -\left( \varphi
_{22}^{A_{3}^{(2)}}*{\cal L}_{2}+2\varphi _{12}^{A_{3}^{(2)}}*{\cal L}%
_{1}\right) (\theta ,r)  \nonumber \\
&&-(g_{12}^{\prime }+g_{23}^{\prime }){\cal L}_{1}(\theta ,r)-g_{22}^{\prime
}{\cal L}_{2}(\theta ,r)\,,\,  \label{TBA22}
\end{eqnarray}
where the kernels are 
\begin{eqnarray}
\varphi _{22}^{A_{3}^{(2)}}(\theta ) &=&\frac{\sin (B\pi /4)}{\cosh
^{2}\theta -\cos ^{2}(B\pi /4)}+\frac{\cos (B\pi /4)}{\cosh ^{2}\theta +\sin
^{2}(B\pi /4)}\,-\frac{2}{\cosh \theta } \\
\varphi _{12}^{A_{3}^{(2)}}(\theta ) &=&2\cosh \theta \left( \frac{2\sin
((B+1)\pi /4)}{\cosh (2\theta )+\sin (B\pi /2)}\,-\frac{\sqrt{2}}{\cosh
(2\theta )}\right) \,.
\end{eqnarray}
Note that $\varphi _{12}^{A_{3}^{(2)}}(\theta )=\varphi
_{23}^{A_{3}^{(2)}}(\theta )$ and $\varphi _{22}^{A_{3}^{(2)}}(\theta
)=\varphi _{11}^{A_{3}^{(2)}}(\theta )+\varphi _{13}^{A_{3}^{(2)}}(\theta )$%
. Alternatively we may write down the universal form of the TBA-equations (%
\ref{uniTBA}), which becomes particularly simple for the fermionic
statistics 
\begin{eqnarray}
\xi _{1}(\theta ,r) &=&(\xi _{2}-L_{2})*\Omega _{4}(\theta ,r)+(L_{1}*\Omega
_{4,B})(\theta ,r)\,\, \\
\xi _{2}(\theta ,r) &=&2(\xi _{1}-L_{1})*\Omega _{4}(\theta
,r)+(L_{2}*\Omega _{4,B})(\theta ,r)\,\,.
\end{eqnarray}
We compute from (\ref{genexp}) or (\ref{funi}) the Fourier transformed
TBA-kernels 
\begin{eqnarray}
\widetilde{\varphi }_{22}^{A_{3}^{(2)}}(k) &=&2\pi \frac{(1+2\cosh (\pi
k/4))(\cosh ((B-1)\pi k/4)-\cosh \pi k/4)}{\cosh (k\pi /4)+\cosh (3k\pi /4)}
\\
\widetilde{\varphi }_{12}^{A_{3}^{(2)}}(k) &=&2\pi \frac{\cosh (k\pi
/4)\cosh ((B-1)k\pi /4)-1}{\cosh (k\pi /2)}\,\,.
\end{eqnarray}
From (\ref{eta2ij}) or from the explicit expression for $\mu _{x}^{(2)}$
corresponding to each block in (\ref{S1}) and (\ref{S22}), we obtain for
fermionic statistics 
\begin{equation}
\eta _{ij}^{(0)}=0\quad \text{and \quad }\eta _{ij}^{(2)}=\frac{\pi
^{2}B(2-B)}{4^{3}}\left( 
\begin{array}{lll}
3 & 2 & 1 \\ 
2 & 4 & 2 \\ 
1 & 2 & 3
\end{array}
\right) _{ij}\,\,.
\end{equation}
Therefore we have 
\begin{equation}
\eta _{1}^{(2)}=\eta _{3}^{(2)}=\frac{3\pi ^{2}B(2-B)}{32}\quad \text{%
and\quad }\eta _{2}^{(2)}=\frac{\pi ^{2}B(2-B)}{8}\,\,.  \label{singeta}
\end{equation}
Computing the sum $\eta _{1}^{(2)}+\eta _{2}^{(2)}+\eta _{3}^{(2)}$ confirms
our general formula (\ref{sumeta}). Finally we obtain for the approximated
scaling function 
\begin{equation}
c(r)=3-\frac{15\pi ^{2}B(2-B)}{16(\beta -\ln (r/2))^{2}}\,\,.
\end{equation}
Once more we compare this analytical expression with the numerical solution
of the two coupled TBA-equations (\ref{TBA12}) and (\ref{TBA22}). Choosing
the constants in the same way as for the Sinh-Gordon model, figure 4
demonstrates the same kind of qualitative agreement as we observed for the
previous models. For all models we observe that at about $r=0.1$ the
analytical expressions for $c(r)$ start to decrease more rapidly than the
numerical solution.

\subsubsection{$(G_{2}^{(1)}\Leftrightarrow D_{4}^{(3)})$-ATFT}

This theory is the simplest example for a model in which the masses as a
function of the coupling constant flow from the classical mass spectrum of
one Lagrangian to its dual in the Lie algebraic sense. The theory contains
two self-conjugate particles whose masses were conjectured \cite{DGZ,KW}
(supported by numerical investigations \cite{WW}) to $m_{1}=m\sin (\pi /H)$
and $m_{2}=m\sin (2\pi /H)$. The ``floating Coxeter number'' is taken to be $%
H=6+3B$ with $0\leq B\leq 2.$ The fusing processes $1\,1\rightarrow 1+2,$ $%
2\,2\rightarrow 1+2+2$ and $1\,2\rightarrow 1+2$ are possible. In our
conventions the related scattering matrices \cite{DGZ,nons} read 
\begin{eqnarray}
&&S_{11}^{G/D}\left( \theta \right) =[1,0]_{\theta }[H-1,0]_{\theta
}[H/2,1/2]_{\theta },\quad S_{12}^{G/D}\left( \theta \right)
=[H/3,1]_{\theta }[2H/3,1]_{\theta },\quad \quad \quad \\
&&S_{22}^{G/D}\left( \theta \right) =[H/3-1,1]_{\theta }[H/3+1,1]_{\theta
}[2H/3-1,1]_{\theta }[2H/3+1,1]_{\theta }\quad .\quad
\end{eqnarray}
Besides the loss of duality these scattering matrices exhibit a further
difference in comparison with the simply laced case, which has a bearing on
the TBA-analysis. In the standard prescription the symmetry of the Bethe
wave function is derived by exploiting the behaviour of the scattering
matrices at $\theta =0$. As already mentioned for the simply laced case we
always have $S_{ij}(0)=(-1)^{\delta _{ij}}$, whereas now we observe $%
S_{22}^{G/D}\left( 0\right) =-S_{11}^{G/D}\left( 0\right) =1$. Assuming that
the particles described are bosons, we should choose, according to the
arguments of \cite{TBAZam1}, $g_{11}=1$ and $g_{22}=0$. In this case,
however, the TBA-analysis does not produce any physical solution. The
TBA-equations in the variant (\ref{eps}) become 
\begin{eqnarray}
\varepsilon _{1}(\theta ,r) &=&rm\sin (\pi /H)\cosh \theta -\left( \varphi
_{11}^{G/D}*{\cal L}_{1}\right) (\theta ,r)-\left( \varphi _{12}^{G/D}*{\cal %
L}_{2}\right) (\theta ,r)\,\,  \nonumber \\
&&-g_{11}^{\prime }{\cal L}_{1}(\theta ,r)-g_{12}^{\prime }{\cal L}%
_{2}(\theta ,r)\,\,, \\
\varepsilon _{2}(\theta ,r) &=&rm\sin (2\pi /H)\cosh \theta -\left( \varphi
_{21}^{G/D}*{\cal L}_{1}\right) (\theta ,r)-\left( \varphi _{22}^{G/D}*{\cal %
L}_{2}\right) (\theta ,r)\,\,  \nonumber \\
&&\,-g_{21}^{\prime }{\cal L}_{1}(\theta ,r)-g_{22}^{\prime }{\cal L}%
_{2}(\theta ,r)\,\,.
\end{eqnarray}
The kernels and their Fourier transformed versions read 
\begin{eqnarray}
\varphi _{11}^{G/D}(\theta ) &=&\omega _{1}(\theta )+\omega _{H-1}(\theta
)+\omega _{H/2,1/2}(\theta ) \\
\varphi _{12}^{G/D}(\theta ) &=&\varphi _{21}(\theta )=\omega
_{H/3,1}(\theta )+\omega _{2H/3,1}(\theta ) \\
\varphi _{22}^{G/D}(\theta ) &=&\omega _{H/3-1,1}(\theta )+\omega
_{H/3+1,1}(\theta )+\omega _{2H/3-1,1}(\theta )+\omega _{2H/3+1,1}(\theta )
\end{eqnarray}
and 
\begin{eqnarray}
\widetilde{\varphi }_{12}^{G/D}(k) &=&4\pi \frac{\cosh ((1+B)\pi k/H)-\cosh
((1-2B)\pi k/H)}{1-2\cosh (\pi k/3)} \\
\widetilde{\varphi }_{11}^{G/D}(k) &=&\frac{\widetilde{\varphi }%
_{12}^{G/D}(k)+4\pi \cosh ((1-B)\pi k/H)}{2\cosh (\pi k/H)} \\
\widetilde{\varphi }_{22}^{G/D}(k) &=&2\cosh (\pi k/H)\widetilde{\varphi }%
_{12}^{G/D}(k)+2\pi \,\,.
\end{eqnarray}
From this or our general formulae of section 4.2. we obtain 
\begin{equation}
\eta _{11}^{(2)}=\frac{\pi ^{2}B(2-B)}{H^{2}}\,,\quad \eta _{12}^{(2)}=-\eta
_{11}^{(2)}\,,\quad \eta _{22}^{(2)}=-2\eta _{11}^{(2)},\quad \eta
_{1}^{(2)}=0,\quad \eta _{2}^{(2)}=-3\eta _{11}^{(2)}\,.
\end{equation}

The facts that $\eta _{1}^{(2)}$ vanishes and $\eta _{2}^{(2)}$ becomes
negative make it impossible to use our approximated analytical solution (\ref
{Soll}). From a numerical point of view not much has changed in comparison
with the previous models and we can still solve the related TBA-equations in
the standard way. The related $L$-functions look qualitatively the same as
in the previous cases and the scaling functions for some values of the
effective coupling are depicted in figure 5 together with the same function
for the $A_{2}^{(1)}$-affine Toda field theory. We observe that for fixed
value of $r$, the scaling function for the $(G_{2}^{(1)},D_{4}^{(3)})$%
-affine Toda field theory is a monotonically increasing function of $B$.
Under the same circumstances the scaling functions related to theories in
which the underlying algebras are simply laced are monotonically decreasing
at first up to the self-dual point $B=1$ and monotonically increase
thereafter.

\vspace{-4cm}
\begin{center}
\includegraphics[width=11cm,height=16cm,angle=-90]{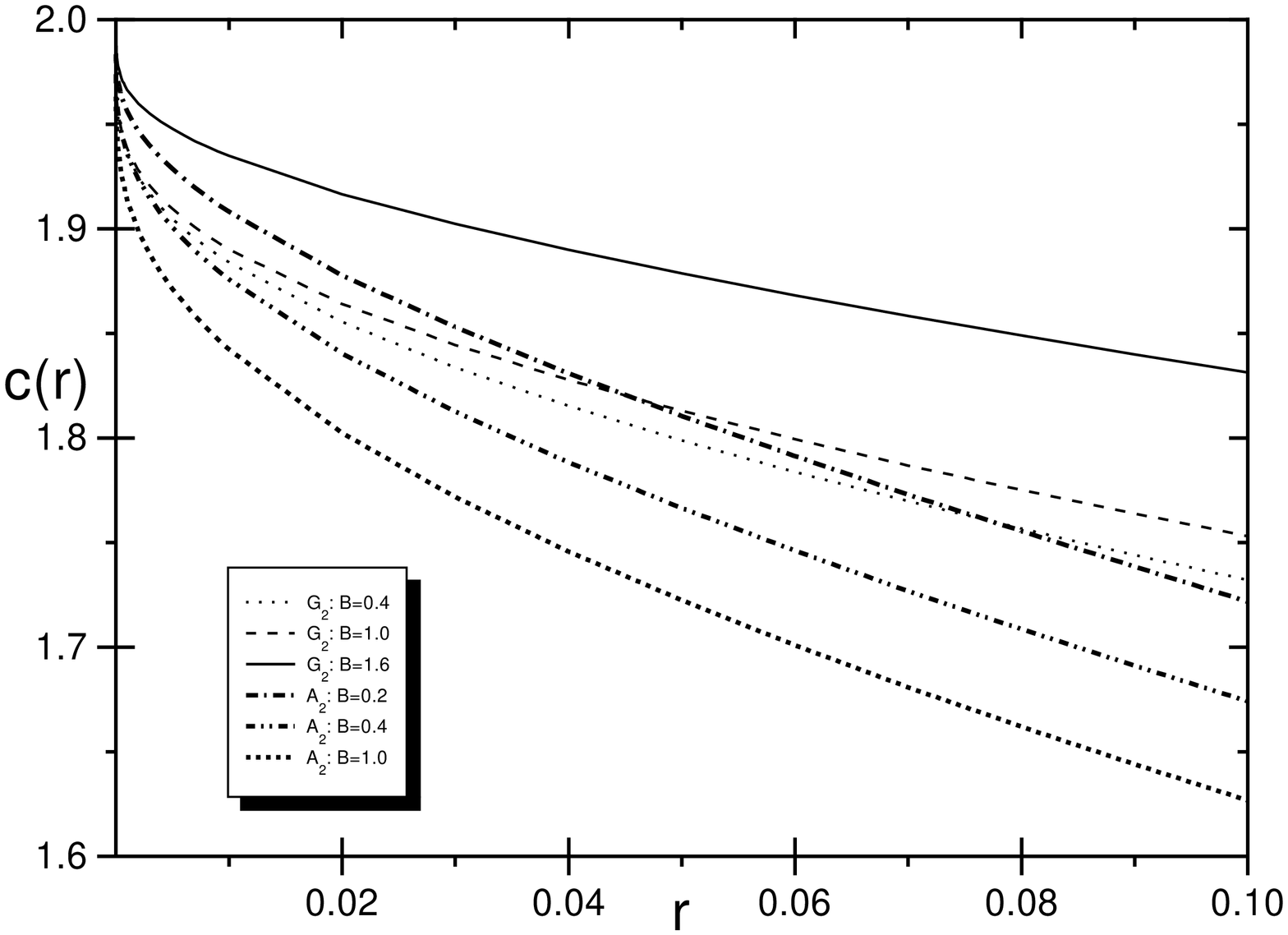}
\end{center}
\vspace{-1cm}
{\small Figure 5: Scaling functions for the
 $A_2^{(1)} $ and $(G_2^{(1)},D_4^{(3)})  $-affine Toda field theory
 related TBA-systems for various values of the effective coupling.}

\section{Existence and Uniqueness Properties}

In this section we are going to investigate the existence and uniqueness
properties of the solutions of the TBA equations. Our main physical
motivation for this considerations is to clarify whether it is possible to
obtain different effective central charges for a fixed dynamical and
statistical interaction due to the existence of several different solutions
of the TBA equation. As a side product we obtain useful estimates on the
error and the rate of convergence of the applied numerical procedure.
Precise estimates of this kind were not obtained previously in this context
and convergence is simply presumed. The method we employ is the contraction
principle (or Banach fixed point theorem), see e.g. \cite{func}. For the
Yang--Yang equation in the case of the non--relativistic one--dimensional
Bose gas for fermionic type of statistics the uniqueness question was
already addressed by Yang and Yang \cite{Yang}, albeit with a different
method.

In order to keep the notation simple we commence our discussion for a system
with one particle only and a statistics of fermionic type. Thereafter we
discuss the straightforward generalization. The standard way to solve
integral equations of the type (\ref{eps}) consists in discretising the
original function and a subsequent iteration, usually numerically.
Normalizing the mass to one, this means we consider (\ref{eps}) as 
\begin{equation}
\varepsilon _{n+1}(\theta ,r):=r\cosh \theta -\varphi *\ln
(1+e^{-\varepsilon _{n}(\theta ,r)})  \label{sequence}
\end{equation}
and perform the iteration starting with $\varepsilon _{0}(\theta ,r)=r\cosh
\theta $. The exact solution is then thought to be the limes $%
\lim_{n\rightarrow \infty }\varepsilon _{n}$. However, a priori it is not
clear whether this limes exists at all and how it depends on the initial
value $\varepsilon _{0}$. In particular, different initial values might lead
to different solutions.

The natural mathematical setup for this type of problem is to re-write the
TBA-equation as 
\begin{equation}
(A\xi )(\theta ,r):=\varphi *\ln (1+e^{\xi (\theta ,r)-r\cosh \theta })=\xi
(\theta ,r),  \label{fixed}
\end{equation}
and treat it as a fixed point problem for the operator $A.$

In order to give meaning to the limes $\lim_{n\rightarrow \infty
}\varepsilon _{n}$ we have to specify a norm. Of course it is natural to
assume that $\xi $ as function of $\theta $ is measurable, continuous and
essentially bounded on the whole real line. The latter assumption is
supported by all known numerical results. Furthermore it follows from (\ref
{fixed}), together with the explicit form of $\varphi $, that possible
solutions $\xi $ vanish at infinity. This means possible solutions of (\ref
{fixed}) constitute a Banach space with respect to the norm

\begin{equation}
\left\| f\right\| _{\infty }=ess\sup \left| f(\theta )\right| ,
\end{equation}
i.e. $L_{\infty }( {\rm I} \!{\rm R})$. 
In principle we are now in a position to apply
the Banach fixed point theorem\footnote{%
One may of course apply different types of fixed point theorems exploiting
different properties of the operator $A$. For instance if $A$ is shown to be
compact one can employ the Leray-Schauder fixed point theorem. In the second
refence of \cite{TBAKM} it is claimed that the problem at hand was treated
in this manner, albeit a proof was not provided.}, which states the following:

Let $D\subset L_{\infty }$ be a nonempty set in a Banach space and let $A$
be an operator which maps $D$ $q$-contractively into itself, i.e. for all $%
f,g\in D$ and some fixed $q$, $0\leq q<1$ 
\begin{equation}
\Vert A(f)-A(g)\Vert _{\infty }\leq q\Vert f-g\Vert _{\infty }.
\label{contraction}
\end{equation}
Then the following statements holds:

\begin{enumerate}
\item[i)]  There exists a unique fixed point $\xi $ in $D$, i.e. equation (%
\ref{fixed}) has exactly one solution.

\item[ii)]  The sequence constructed in (\ref{sequence}) by iteration
converges to the solution of (\ref{fixed}).

\item[iii)]  The  error of the iterative procedure may be estimated by
\[
\Vert \xi -\xi_{n}\Vert _{\infty }\leq \frac{q^{n}}{1-q}\Vert 
\xi_{1}-\xi_{0}\Vert
_{\infty }\quad \text{and\quad }\left\| \xi -\xi_{n+1}\right\| _{\infty }\leq 
\frac{q}{1-q}\left\| \xi_{n+1}-\xi_{n}\right\| _{\infty }.
\]

\item[iv)]  The rate of convergence is determined by 
\[
\left\| \xi -\xi_{n+1}\right\| _{\infty }\leq q\left\| \xi -\xi_{n}\right\|
_{\infty }.
\]
\end{enumerate}

In order to be able to apply the theorem we first have to choose a suitable
set in the Banach space. We choose some $q\in [0,1)$ such that $e^{-r}\leq q$
and take $D$ to be the convex\footnote{%
For $f,g\in D_{q,r}$ also $tf+(1-t)g\in D_{q,r}$, with $0\leq t\leq 1$.} set
$D_{q,r}:=\left\{ f:\text{ }\Vert f\Vert _{\infty }\leq \ln \frac{q}{1-q}%
+r\right\} $. We may now apply the following estimate for the convolution
operator $\varphi *$ (which is a special case of Young's inequality) 
\begin{equation}
\Vert \varphi *f\Vert _{\infty }\leq \Vert \varphi \Vert _{1}\Vert f\Vert
_{\infty }.  \label{convolution}
\end{equation}
For the concrete one particle models at hand, the Sinh-Gordon- and the
Bullough-Dodd-model, we have $\Vert \varphi \Vert _{1}:=\int \frac{d\theta }{%
2\pi }|\varphi (\theta )|=1$. For the operator $L$ we have the estimate 
\[
\medskip L(f)\leq \ln \left[ 1+\exp \left( \left\| f\right\| _{\infty
}-r\right) \right] \leq \ln \frac{1}{1-q}\leq \ln \frac{q}{1-q}+r.
\]
The last inequality follows from our special choice of $q$. Thus, $A$ maps $%
D_{q,r\text{ }}$into itself.

In the final  step we show that the contraction property (\ref{contraction})
is fulfilled on $D_{q,r\text{ }}$. It suffices to prove this for the map $L$%
, because of (\ref{convolution}) and the fact that $\Vert \varphi \Vert
_{1}=1$. We have  
\begin{eqnarray*}
\Vert L(f)-L(g)\Vert _{\infty } &=&\left\| \tint_{0}^{1}dt\,\frac{d}{dt}%
L(g+t(f-g))\right\| _{\infty } \\
&=&\left\| \tint_{0}^{1}dt\,\frac{(f-g)}{1+\exp (-g-t(f-g)+r\cosh \theta )}%
\right\| _{\infty } \\
&\leq &\max_{0\leq t\leq 1}\left| \frac{1}{1+\exp (-g-t(f-g)+r)}\right|
\left\| f-g\right\| _{\infty } \\
&\leq &\max_{0\leq t\leq 1}\left| \frac{1}{1+\exp \left( -\left\|
g-t(f-g)\right\| _{\infty }+r\right) }\right| \left\| f-g\right\| _{\infty }
\\
&\leq &q\left\| f-g\right\| _{\infty }\,\,.
\end{eqnarray*}
In the last inequality we used the fact that $D_{q,r\text{ }}$ is a convex
set.

We may now safely apply the fixed point theorem. First of all we conclude
from i) and ii) that a solution of (\ref{fixed}) not only exists, but it is
also unique. In addition we can use iii)and iv) as a criterium for error
estimates. From our special choice of the closed set $D_{q,r}$ one sees that
the rate of convergence depends crucially on the parameter $r$, the smaller $%
r$ the greater $q$ is, whence the sequence $(\xi_{n})$ converges slower.

One could be mathematically more pedantic at this point and think about 
different requirements on the function $\xi$. For instance one might allow 
functions which are not bounded (we do not know of any example except
when $r \rightarrow 0 $) and then pursue similar arguments as before
on $L_p$ rather than $ L_{\infty }$.   

The generalization of the presented arguments to a situation involving $l$
different types of particles and general Haldane statistics may be carried
out in a straightforward manner. 

\section{Conclusions}

Clearly an unsatisfactory feature of our approximated analytical expression
for the scaling function (\ref{universal}) is that, at the moment, we do not
have any additional constraint at hand which allows to determine the
constant $\beta $. Admittedly the choice (\ref{Konstanten}) enters our
analysis in a rather ad hoc way. At the moment the rational behind our
choice is that we may compare it with the semi-classical results in the
spirit of \cite{ZamoR} and thereafter restore the strong-weak duality. In
addition it is supported by our numerical results. Surely this is by no
means unique and it is highly desirable to eliminate this ambiguity.

Nonetheless, different choices of the constant will not change the overall
behaviour and our expression is clearly in conflict with the results of
Cassi and Destri \cite{Cassi}, who found $f(r)=-3r^{2}m_{1}^{2}/(\pi \varphi
_{11}^{(1)})$ for the function in equation (\ref{typical}). The convention
therein are that $1$ labels the particle type with smallest mass and $%
\varphi _{11}^{(1)}$ results from the power series $\varphi _{11}(\theta
)=-\sum_{k=1}^{\infty }\varphi _{11}^{(k)}\exp (-k\left| \theta \right| )$.
On the side of the TBA-analysis the origin of this discrepancy may be
tracked down easily. A behaviour of the type quoted in \cite{Cassi} was
derived before in \cite{ZamoR,TBAKM} for affine Toda field theories related
to the minimal part of the scattering matrix (\ref{S}). This result was then
extrapolated by Cassi and Destri in the way that $\varphi _{11}^{(1)}$ was
taken to be the coefficient of the power series related to the full coupling
constant dependent scattering matrix. However, the derivations in \cite
{ZamoR,TBAKM} rely heavily on the assumption that the function $L(\theta )$
is constant in the region $-\ln (2/r)\ll \theta \ll \ln (2/r)$. It is
essentially this property which is lost for the full theory (as our
numerical results demonstrate), such that the arguments of Cassi and Destri
become faulty. It would be interesting to settle the question also on the
perturbative side and bring our results into agreement with perturbation
theory. Our findings will surely have consequences for the subtraction
scheme used in such considerations.

There exist of course other methods which allow to extract the ultraviolet
central charge from a massive integrable model. The c-theorem \cite{ZamoC}
has turned out to be very efficient in this context and a direct comparison
is very suggestive. As the main input, the c-theorem requires the n-particle
form factors (the 2-particle form factor is usually sufficient) related to
the trace of the energy-momentum tensor. Unlike in the situation of
conformal invariance, in which the trace vanishes, this tensor is not unique
and acquires some scaling behaviour for massive theories. As observed in 
\cite{MS}, the consequence of this fact is that the c-theorem produces a
whole ray of central charges greater than the rank of the underlying Lie
algebra ${\bf g}$. The natural question which arises is to identify the
origin of this ambiguity inside the TBA-approach. One might assume that it
results from several possible solutions of the TBA-equation. However, this
possibility is definitely ruled out as follows from our investigations in
section 5. Since the statistical interaction $g$ enters the analysis as a
further parameter, this could provide a further possible mechanism which
produces the observed values. As our investigations demonstrate, however,
also this possibility can be excluded for certain, since different choices
for $g$ only produce central charges smaller than the rank of ${\bf g}$. In
the light of this results we conjecture that the responsible mechanism is to
include a non-vanishing chemical potential in the way as was already
indicated by Yang and Yang \cite{Yang}. To settle this question precisely
requires more detailed investigations \cite{FKS}.

There are several further questions which should be addressed in order to
complete the picture. It will certainly be interesting to obtain also the
higher terms beyond the first leading order in (\ref{typical}) and to
exploit further the Y-systems of section 4.4. as an alternative analytical
approach. There exist interesting links between these systems and spectral
functions \cite{spectral}, such that one can expect more exact and universal
results to follow.

\noindent {\bf Acknowledgments: } We would like to thank M.~Schmidt for
useful discussions and comments, C.~Figueira~de~Morisson~Faria for
help with the figures and T. Oota for drawing our attention to ref. 
\cite{Oota}. A.F. and C.K. are grateful to the Deutsche
Forschungsgemeinschaft (Sfb288) for partial financial support.

\begin{description}
\item  {\small \setlength{\baselineskip}{12pt}}
\end{description}

\end{document}